\documentclass[sn-basic]{sn-jnl}

\jyear{2021}%

\theoremstyle{thmstyleone}%
\theoremstyle{thmstyletwo}%

\theoremstyle{thmstylethree}%

\usepackage{subfig}
\begin{document}

\title[Five-year in-orbit background of \textit{Insight}-HXMT]{Five-year in-orbit background of \textit{Insight}-HXMT}


\author*[1]{\fnm{Jin-Yuan} \sur{Liao}}\email{liaojinyuan@ihep.ac.cn}

\author[1]{\fnm{Shu} \sur{Zhang}}     

\author[1]{\fnm{Juan} \sur{Zhang}}    
\author[1]{\fnm{Gang} \sur{Li}}       

\author[1]{\fnm{Zhi} \sur{Chang}}     
\author[1]{\fnm{Yu-Peng} \sur{Chen}}  
\author[1]{\fnm{Ming-Yu} \sur{Ge}}    
\author[1]{\fnm{Jing} \sur{Jin}}      
\author[1]{\fnm{Xue-Feng} \sur{Lu}}   
\author[1]{\fnm{Yuan} \sur{You}}      

\author[1]{\fnm{Xue-Lei} \sur{Cao}}    
\author[1]{\fnm{Yong} \sur{Chen}}    
\author[1]{\fnm{Yue} \sur{Huang}}       
\author[1]{\fnm{Wei-Chun} \sur{Jiang}} 
\author[1]{\fnm{Xiao-Bo} \sur{Li}}      
\author[1]{\fnm{Xu-Fang}   \sur{Li}}     
\author[1]{\fnm{Zheng-Wei} \sur{Li}}     
\author[1]{\fnm{Cong-Zhan} \sur{Liu}}    
\author[1]{\fnm{Ying}    \sur{Tan}}    
\author[1]{\fnm{Yan-Ji} \sur{Yang}}  
\author[1]{\fnm{Yi-Fei}    \sur{Zhang}}  
\author[1]{\fnm{Hai-Sheng} \sur{Zhao}}  

\author[1,2]{\fnm{Fang-Jun} \sur{Lu}}
\author[1]{\fnm{Yu-Peng} \sur{Xu}}
\author[1]{\fnm{Jin-Lu} \sur{Qu}}
\author[1]{\fnm{Li-Ming} \sur{Song}}
\author[1,3,4]{\fnm{Shuang-Nan} \sur{Zhang}}

\affil*[1]{\orgdiv{Key Laboratory of Particle Astrophysics}, \orgname{Institute of High Energy Physics, Chinese Academy of Sciences}, 
\orgaddress{\street{19B Yuquan Road, Shijingshan District}, \city{Beijing}, \postcode{100049}, \country{China}}}

\affil[2]{\orgdiv{Key Laboratory of Stellar and Interstellar Physics and School of Physics and Optoelectronics}, \orgname{Xiangtan University}, \orgaddress{\street{Yuhu District}, \city{Xiangtan}, \postcode{411105}, \country{China}}}

\affil[3]{\orgname{University of Chinese Academy of Sciences}, \orgaddress{\street{19A Yuquan Road, Shijingshan District}, \city{Beijing}, \postcode{100049}, \country{China}}}

\affil[4]{\orgdiv{Key Laboratory of Space Astronomy and Technology}, \orgname{National Astronomical Observatories, Chinese Academy of Sciences}, \orgaddress{\street{20A Datun Road, Chaoyang District}, \city{Beijing}, \postcode{100012}, \country{China}}}




 \abstract{
 \textbf{Purpose:} We present the five-year in-orbit background evolution of \textit{Insight}-HXMT since the launch, as well as the effects of the background model in data analysis.
 \textbf{Methods:} The backgrounds of the three main payloads, i.e., Low-Energy Telescope, Medium-Energy Telescope and High-Energy Telescope, are described, respectively.
 The evolution of the background over time is obtained by simply comparing the background in every year during the in-orbit operation of \textit{Insight}-HXMT.  \textbf{Results:} The major observational characteristics of the \textit{Insight}-HXMT in-orbit background are presented, including the light curve, spectrum, geographical distribution, and long-term evolution. The systematic error in background estimation is investigated for every year.
 \textbf{Conclusion:} The observational characteristics of the five-year in-orbit background are consistent with our knowledge of the satellite design and the space environment, and the background model is still valid for the latest observations of \textit{Insight}-HXMT.
 }




\maketitle
\section{Introduction}\label{sec:intro}

\begin{figure}[ht]%
	\centering
		\includegraphics[width=0.6\textwidth]{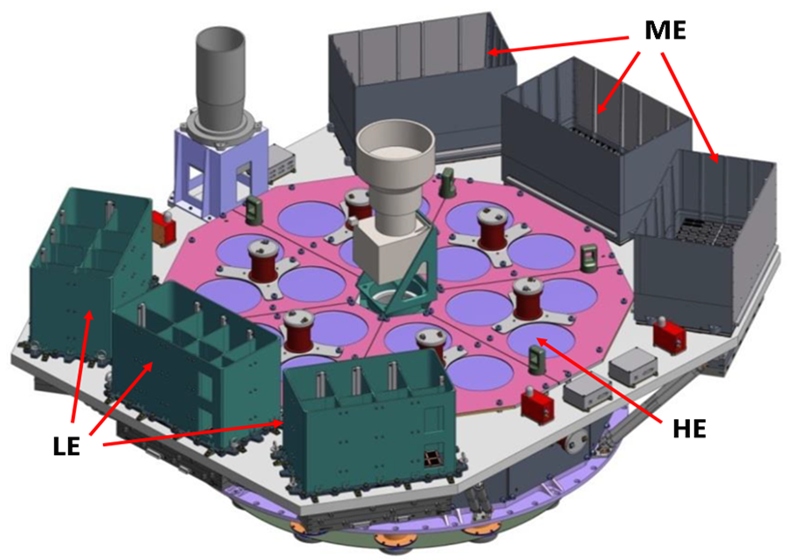}
		\caption{Main structure of \textit{Insight}-HXMT.}
		\label{fig:HXMT}
\end{figure}

Up to June 2022, the Hard X-ray Modulate Telescope (dubbed as \textit{Insight}-HXMT), the first space X-ray telescope of China, has been operated successfully in orbit for five years since the launch on June 15, 2017 \citep{ZSN2020}. There are three main payloads onboard \textit{Insight}-HXMT, namely, Low Energy Telescope (LE; \citealt{ChenY2020}), Medium Energy Telescope (ME; \citealt{CaoXL2020}) and High Energy Telescope (HE; \citealt{LiuCZ2020}). All three payloads working in the collimated way are combined to cover a broad energy range from 1 keV to 250 keV. The layout of the three telescopes on the main structure is shown in Fig.~\ref{fig:HXMT}, and the main parameters are shown in Table~\ref{tab:para}.

There are three main observation modes assigned to \textit{Insight}-HXMT: pointing observation, Galactic plane scanning survey and gamma-ray burst observation.
Although the data analysis methods and processes are different, the background estimation is one of the crucial steps for fulfilling each task. For gamma-ray burst observation, the background is usually taken from pre- and post-burst \citep{Luo2020}. For scanning observation, the background is derived from the scanning light curves with sources subtracted \citep{Sai2020}. However, the background estimation of the pointing observation is much more complicated, since the traditional on-off mode of \textit{BeppoSAX}/PDS \citep{Frontera1997a,Frontera1997b} and \textit{RXTE}/HEXTE \citep{Rothschild1998} for observation of the background was not adopted in the design of \textit{Insight}-HXMT. 
Therefore, in order to estimate the background accurately, a dedicated background model has been constructed during the operation of \textit{Insight}-HXMT \citep{Liao2020_le,Liao2020_he,Guo2020}.

Before the launch of \textit{Insight}-HXMT, thorough background simulation works have been done (e.g., \citealt{LiG2009,XieF2015}). With the Geant4 tools and the mass model of \textit{Insight}-HXMT, these works result in the expected in-orbit background of \textit{Insight}-HXMT, such as the background spectra induced by various particles. The first two-year background observation of \textit{Insight}-HXMT proved the consistency between the in-orbit observation and the on-ground simulation, which also suggests that our understanding of the in-orbit background of \textit{Insight}-HXMT is reliable \citep{ZhangJ2020}.

\begin{figure}[ht]%
	\centering
		\includegraphics[width=0.9\textwidth]{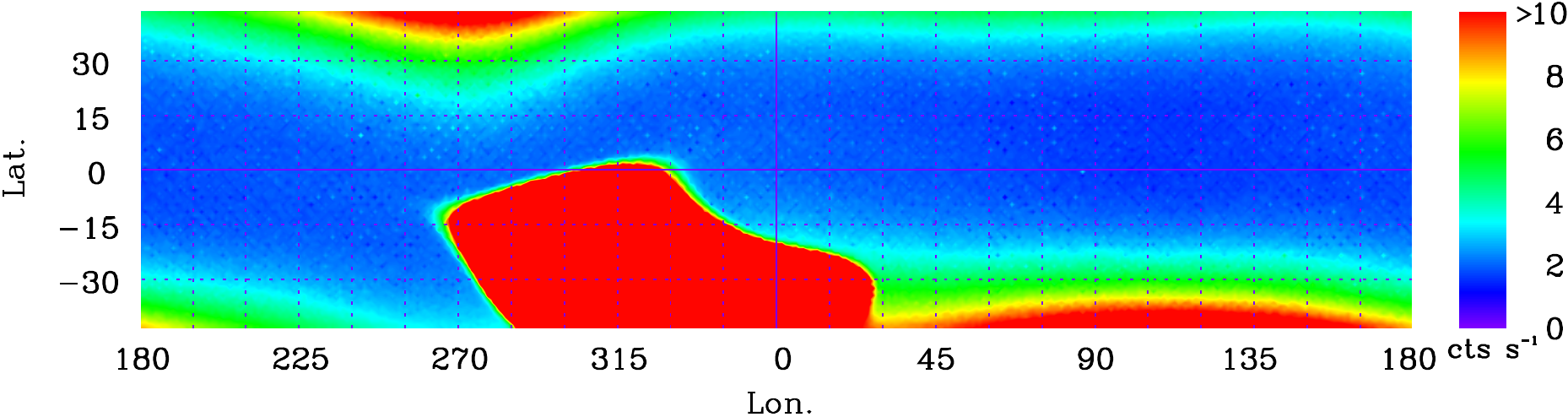}
		\includegraphics[width=0.9\textwidth]{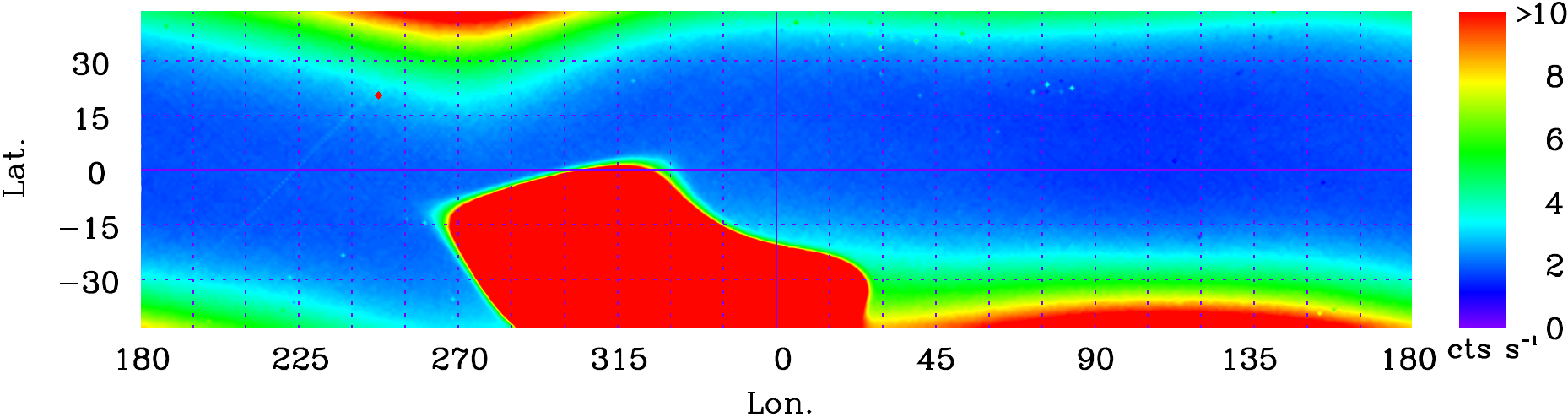}
		\caption{Geographical distributions of the PM count rate in the first {\bf (top)} and fifth {\bf (bottom)} years.}
		\label{fig:PM_map}
\end{figure}

\textit{Insight}-HXMT is a low Earth orbit satellite. The orbit is approximately a circle with an attitude 550~km and an inclination 43$^\circ$. As shown in previous works (e.g., \citealt{LiG2009,XieF2015,Liao2020_le,Liao2020_he,Guo2020,ZhangJ2020}), the orbital space environment of \textit{Insight}-HXMT is complex and various particles can interact with the satellite platform and instruments to generate many background components (e.g., \citealt{Alcaraz2000a,Alcaraz2000b}). The cosmic-ray protons contribute most to the background of \textit{Insight}-HXMT, while the electrons, neutrons, cosmic X-ray background (CXB), as well as earth albedo of gamma-rays can also contribute to the background.
\textit{Insight}-HXMT can explore the environment of charged particles with its Particle Monitor (PM; \citealt{LuXF2020}), which is on board in the top panel of the satellite platform and sensitive to protons ($>20$~MeV) and electrons ($>1.5$~MeV). As shown in Fig.~\ref{fig:PM_map}, the geographical distribution of the PM count rate has varied little over the past five years for the whole orbit of \textit{Insight}-HXMT, including the low-latitude and high-latitude region, as well as the South Atlantic Anomaly (SAA). For the region with low count rate near the magnetic equator, the average count rate is stable at 1.7--1.8~$\rm cts~s^{-1}$. Therefore, the space environment has changed very little during the last five years. However, we have observed some long-term variations in backgrounds, which are largely related to some degradation effects for LE and ME detectors and  the activation effects of HE by bombardment of charged particles.
Therefore, the background of \textit{Insight}-HXMT after five years of operation needs to be reviewed, including the evolution of the observed characteristics and the validity of the background model.
In this paper, the blank sky observations in the high Galactic latitude region (Table~\ref{tab:obsid}) during the five years since the launch of \textit{Insight}-HXMT are used to investigate the in-orbit background of the three main payloads of \textit{Insight}-HXMT, including the observational characteristics and the systematic error analysis of the background model.

\begin{figure}[ht]%
	\centering
		\includegraphics[width=0.3\textwidth]{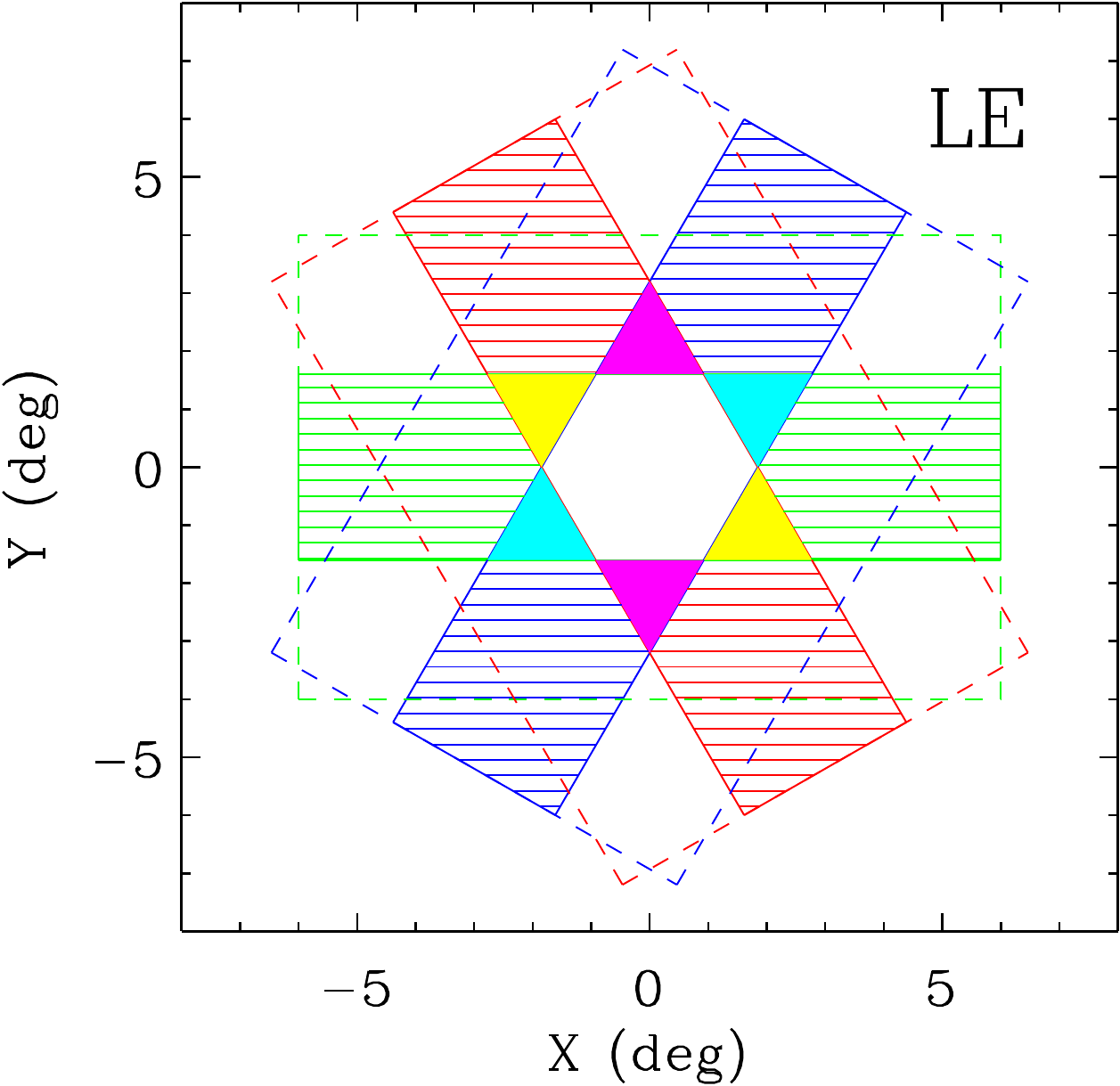}
		\hspace{.1in}
		\includegraphics[width=0.3\textwidth]{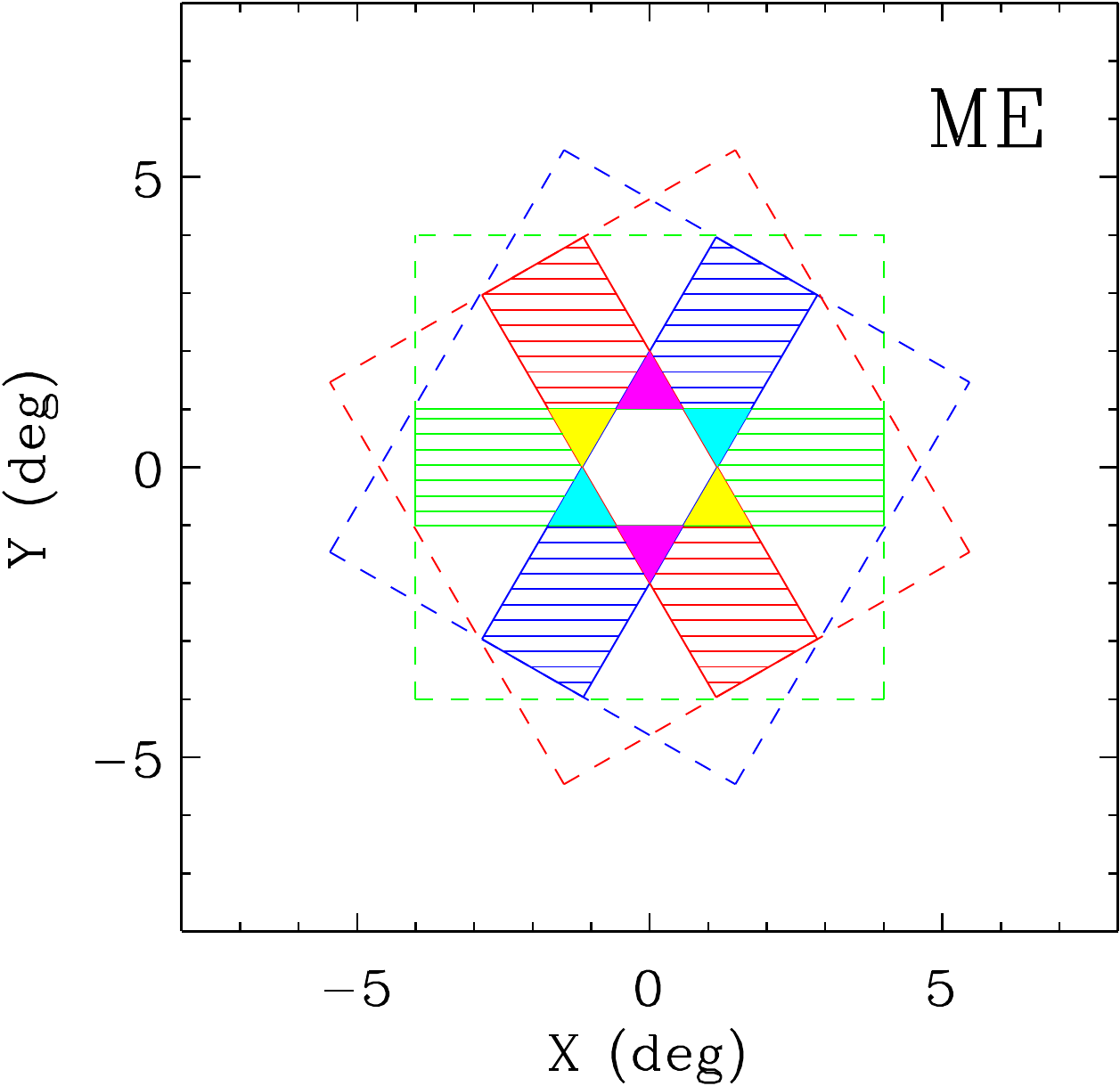}
		\hspace{.1in}
        \includegraphics[width=0.3\textwidth]{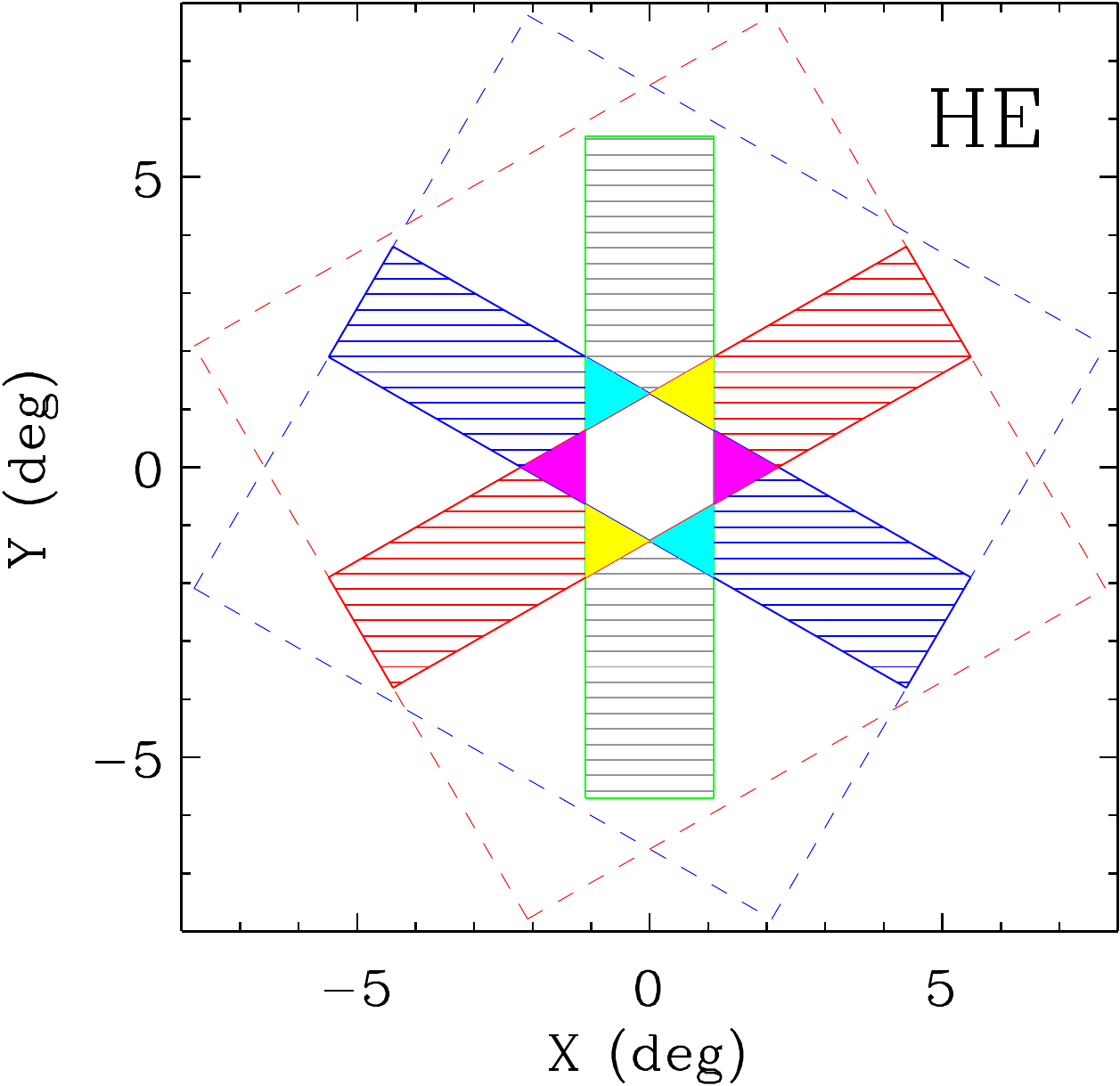}
		\caption{FoVs of LE, ME and HE.}
		\label{fig:FOV}
\end{figure}

It is worth noting that all the three telescopes of \textit{Insight}-HXMT have various Field of Views (FoVs) with different sizes and orientations (Fig~\ref{fig:FOV} and Table~\ref{tab:para}). For the pointing observations of LE and ME, the small FoV detectors are encouraged to be used for scientific analysis. Accordingly, the observed characteristics and background models of LE and ME in this paper mainly focus on these detectors.
This paper is organized as follows. The backgrounds of LE, ME and HE are described in Sections 2--4, respectively. In Section 5, a summary and conclusion are given. 

\begin{table}[ht]
\begin{center}
\begin{minipage}{260pt}
\caption{Main instrumental parameters of LE, ME and HE.}\label{tab:para}%
\begin{tabular}{@{}llll@{}}
\toprule
   & LE  & ME & HE\\
\midrule
Detector Type       &  SCD      & Si-PIN  &  Phoswich  \\
Energy range (keV)\footnotemark[1]  &  0.7--13  & 5--40   & 20--250  \\
Geometrical area (cm$^2$)  &  384  &  952  &  5096  \\
Small FoV (FWHM) & $1.6^\circ\times6^\circ$ & $1^\circ\times4^\circ$ & $1.1^\circ\times5.7^\circ$  \\
Large FoV (FWHM) & $4^\circ\times6^\circ$   & $4^\circ\times4^\circ$ & $5.7^\circ\times5.7^\circ$  \\
\botrule
\end{tabular}
\footnotetext[1]{Design value that will be adjusted over time.}
\end{minipage}
\end{center}
\end{table}

\begin{table}[ht]
\begin{center}
\begin{minipage}{280pt}
\caption{Observation of \textit{Insight}-HXMT background in five years}\label{tab:obsid}%
\begin{tabular}{@{}lll@{}}
\toprule
  ObsID  &  Duration  &  Target\footnotemark[1]  \\
\midrule
P0101293 (001-191)  &  2017-11-02 to 2019-06-26  &  Blank Sky  \\
P0202041 (001-161)  &  2019-07-10 to 2020-07-22  &  Blank Sky  \\
P0301293 (001-115)  &  2020-08-06 to 2021-08-30  &  Blank Sky  \\
P0401293 (001-115)  &  2021-09-14 to 2022-08-29  &  Blank Sky  \\
\midrule
P0101297 (201-217)  &  2017-09-13 to 2018-09-14  &  PSR B0540-69  \\
P0101322 (001-001)  &  2017-07-19 to 2017-07-23  &  PSR B0540-69  \\
P0114550 (001-003)  &  2017-09-20 to 2017-09-27  &  GW~170817  \\
P0101326 (001-018)  &  2017-07-08 to 2019-02-19  &  Cas~A  \\
P0202041 (200-208)  &  2019-07-13 to 2020-07-29  &  Cas~A  \\
P0302291 (001-020)  &  2020-08-23 to 2021-08-21  &  Cas~A  \\
P0101326 (001-015)  &  2021-09-17 to 2022-08-19  &  Cas~A  \\
\botrule
\end{tabular}
\footnotetext[1]{The blank sky observations can be used as the background observations of LE, ME and HE. The observations of PSR~B0540-69, GW~170817 and Cas~A are used only for HE background analysis as the sources are too weak to be detected by HE, and thus can be regarded as `Blank sky'.}
\end{minipage}
\end{center}
\end{table}

\section{Background of the Low-Energy Telescope}
\label{sec:le_bkg}

LE is made of Swept Charge Device (SCD) detectors with a geometrical area of 384~${\rm cm^2}$ and a band pass of 1--13~keV. 
LE has three boxes with the FoV orientation differing by 60$^\circ$. Each box has 20 small FoV detectors (some are broken one after another), 6 large FoV detectors, 2 detectors with their collimators blocked by aluminum covers.
The blocked small FoV detector is designed to measure the particle background, and the blocked large FoV detector is planted with a $^{55}$Fe radioisotope to monitor the energy response.
During the five years operation of \textit{Insight}-HXMT, some of the LE detectors were broken and shut down. The details of the LE bad detectors can be obtained from the `Bad Detector FITS file' that is included in the \textit{Insight}-HXMT Data Analysis software.

Fig.~\ref{fig:le_lc} presents the LE light curve of a blank sky, which shows the typical outline of the LE background embedded with a series of special features.
The whole time range can be divided into the abnormal and normal stages of the instrument.
For the abnormal stage, LE is usually troubled by a large amount of low-energy charged particles and visible light due to the relatively large FoV, which enter from the collimator and are difficult to accurately estimate. In severe cases, LE detectors will be saturated with in-orbit storage overflows.
The instrument's normal stage can be divided into three types. First is the time interval of earth occlusion, where the light curves of the detectors with different FoVs coincide with each other and no CXB photons are recorded. Second is the flare time interval and the flares can be detected in both the small and large FoV detectors. Moreover, the flare flux is basically proportional to the FoV size. Finally, the time interval with neither earth occlusion nor flare is considered as the good time interval (GTI). The usual scientific analysis only uses the GTI data. In order to estimate the background accurately, the background analysis procedure deals with both the regular GTI judgement, and its count rate comparison between the small and large FoV detectors \citep{Liao2020_le}. 
The observational characteristics of the LE background and the effectiveness of the background model in the past five years are shown in what follows.

\subsection{Observational characteristic and long-term evolution of the LE background}
\label{sec:le_obs_char}
\begin{figure}[ht]%
\centering
\includegraphics[width=0.6\textwidth]{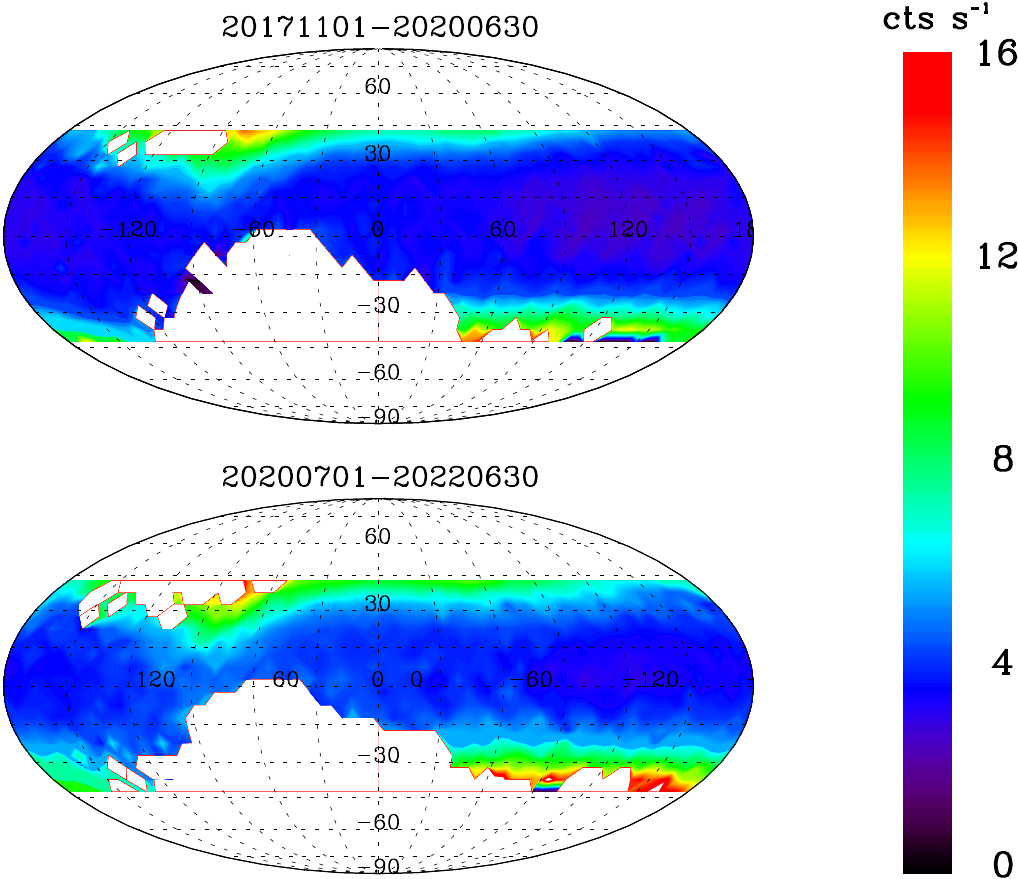}
\caption{Geographical distributions of the background of LE small FoV detectors before (top) and after (bottom) 2020-06-30.}\label{fig:le_map}
\end{figure}

\begin{figure}[ht]%
	\centering
		\includegraphics[width=0.6\textwidth]{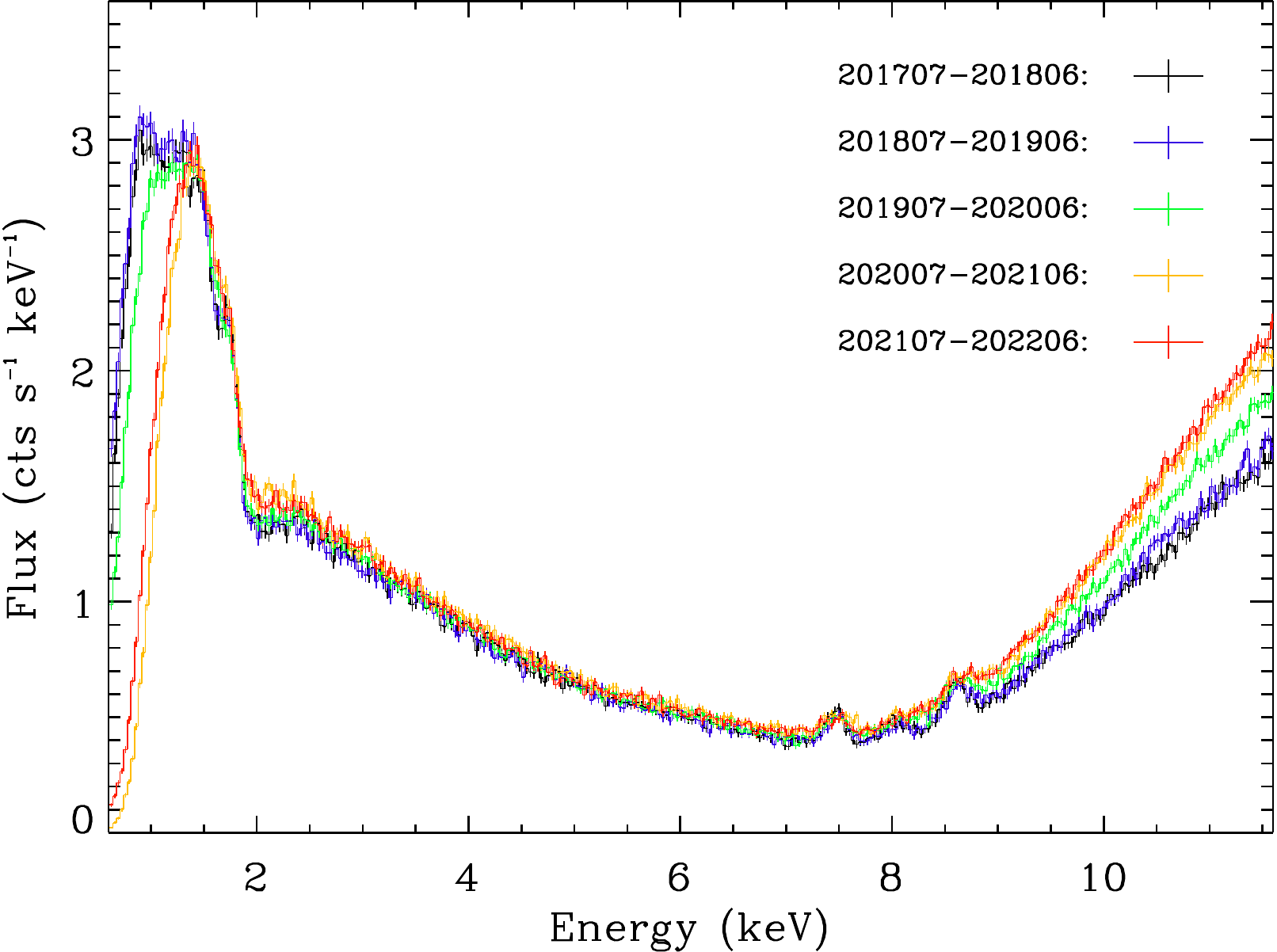}
		\caption{Spectra of the LE small FoV detectors for every year since \textit{Insight}-HXMT operate in orbit.}
		\label{fig:le_spec_small}
\end{figure}

\begin{figure}[ht]%
	\centering
		\includegraphics[width=0.6\textwidth]{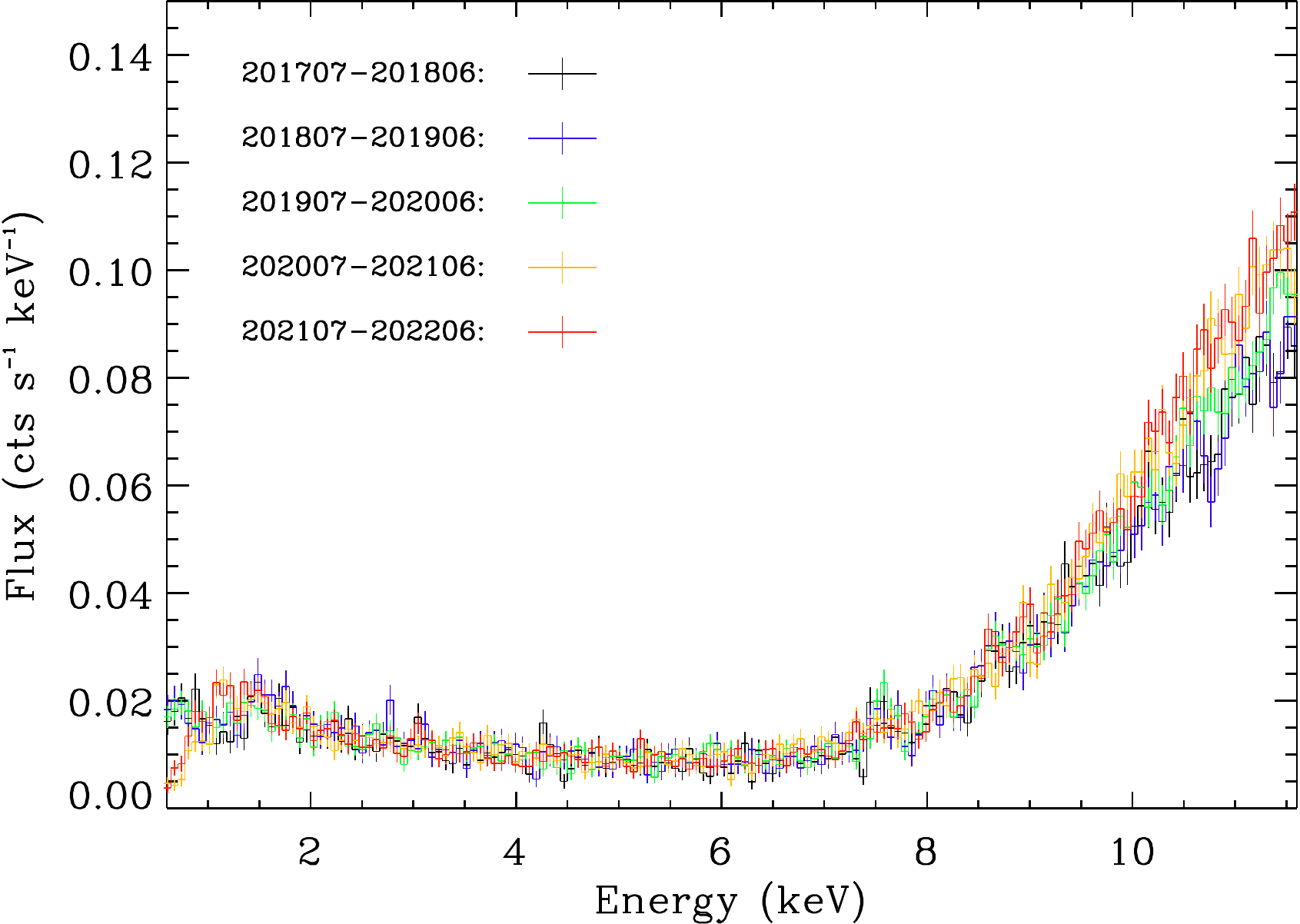}
		\caption{Same as Fig.~\ref{fig:le_spec_small} but for the LE blocked FoV detectors.}\label{fig:le_spec_blind}
\end{figure}

Fig.~\ref{fig:le_map} shows the comparison of the geometrical distributions of the LE background before and after 2021-06-30. It can be seen that the distribution with longitude ($lon$) and latitude ($lat$) has not changed, but the intensity has increased significantly.
The spectra of the same geographical region ($55^\circ<lon<210^\circ$, $-15^\circ<lat<15^\circ$) are specifically investigated for every year. Fig.~\ref{fig:le_spec_small} shows the 5-year background spectra of the small FoV detectors. It can be seen that the spectra have little change at the low energy end, while showing a gradually increasing trend at the high energy end. The continuous broadening of the emission lines is the result of the decline of the LE energy resolution. As demonstrated by the on-ground simulations \citep{ZhangJ2020} and previous in-orbit observations, the LE background can be simplified into a diffuse X-ray background dominant in low energy band and a particle background dominant in high energy band. Thus, the difference between the two geographical distributions shown in Fig.~\ref{fig:le_map} is mainly due to the change in the high energy band. Fig.~\ref{fig:le_spec_blind} shows the 5-year background spectra of the blocked FoV detectors, and the results are consistent with these of the small FoV detectors.
As described in \cite{ZhangJ2020}, the background of \textit{Insight}-HXMT can be produced by various incidence. The background that can be recorded immediately after the incidence is called the prompt background, and the background recorded for a long time (hour to month) after the incidence is called delayed background. It is worth noting that both the backgrounds caused by the CXB and cosmic-ray protons are prompt background.

\begin{figure}[ht]%
\centering
\includegraphics[width=0.6\textwidth]{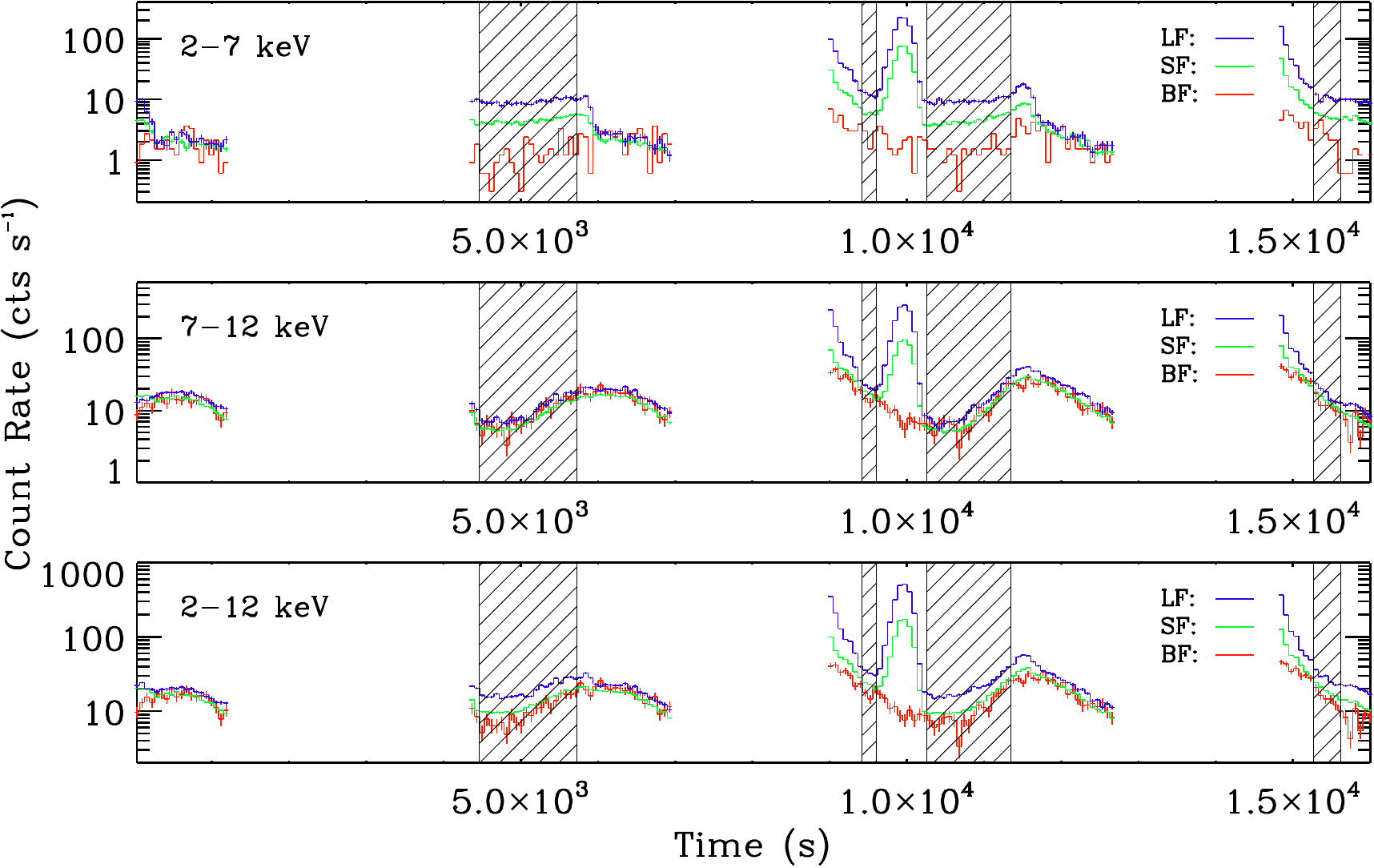}
\caption{Light curve of the LE small (green), big (blue) and blocked (red) FoV detectors (ObsID: P030129303601).}\label{fig:le_lc}
\end{figure}

The behaviour of the LE background light curve has changed very little over the five years. The most obvious features are that the count rate is stable in the low energy band and is modulated significantly by the geomagnetic field in the high energy band, which is also shown in Fig.~\ref{fig:le_lc}.

\subsection{Validity of the LE background model}
\label{le_bkg_model}

\cite{Liao2020_le} has found that the background spectral shape of LE blocked FoV detectors does not change with geographical location, and can be used to characterize the particle background spectral shape of small FoV detectors. The LE background model just takes advantage of this feature to give a simple and reliable background estimation.

There is an evolution of the LE background over the five years, although it is not very significant (Fig.~\ref{fig:le_map}). 
It can be seen from Fig.~\ref{fig:le_spec_small}--\ref{fig:le_spec_blind} that the small and blocked FoV detectors have a similar evolution trend, i.e., the lower limit of the spectral energy range becomes higher and the count rate becomes larger as the in-orbit time increases.
For the LE detector, a large signal can be recorded as several split-events in several pixels at the same time. However, only the events above a certain threshold will be recorded and involved in the subsequent split-event reconstruction. For example, a large signal with energy E can be recorded as two signals with E0 and E1 ($\rm E0+E1=E$). If E1 is less than the threshold, this large signal will be treated as a single-event with E0. With increasing irradiation damage of the LE detector, the distribution of noise signal becomes wider. In order to eliminate the influence of noise signal on the working energy band, the threshold is also adjusted higher. This will raise the lower energy limit of the LE detector, as shown in the low energy band of the spectra in Fig.~\ref{fig:le_spec_small}--\ref{fig:le_spec_blind}. Moreover, the small signals that can be recorded and participate in the split-event reconstruction before threshold adjustment will not exceed the threshold after that, i.e., the double-events that can be reconstructed before will not be reconstructed after threshold adjustment. As the threshold becomes higher, a larger proportion of the double-events will not be reconstructed but will be treated as single-events with a lower energy. As shown in Fig.~\ref{fig:le_spec_small}--\ref{fig:le_spec_blind} for the evolution of the background spectra, the background spectrum shifts to the left year by year. Therefore, the increasing trend of LE background is the result of LE detector irradiation damage and split-event reconstruction in data processing.
Moreover, the spectra of LE blocked FoV detectors in high energy band can be mixed with the super-threshold signal peaks, so the energy band of the blocked FoV detectors in background model has been adjusted accordingly.

\begin{figure}[ht]%
	\centering
		\includegraphics[width=0.6\textwidth]{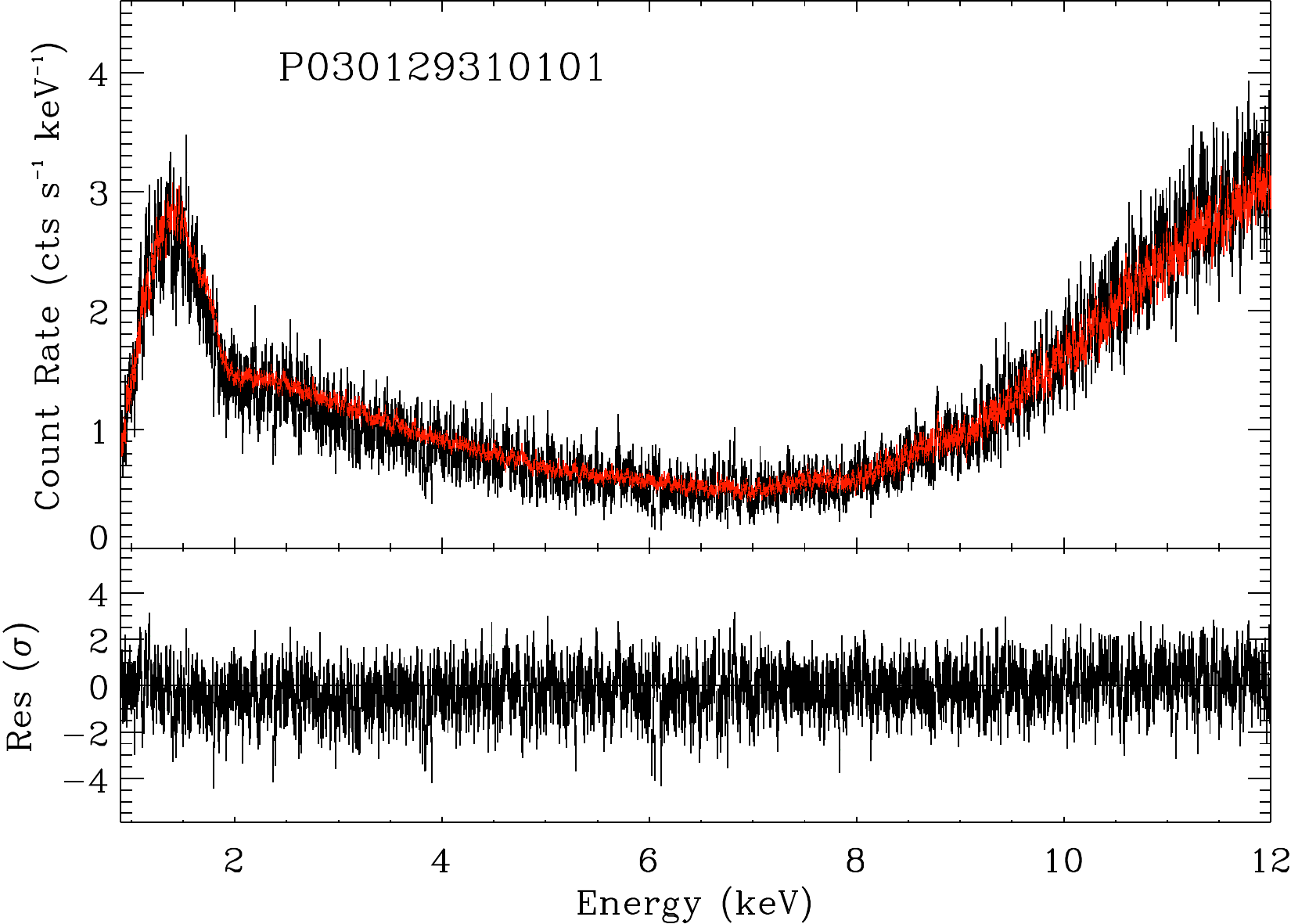}
		\caption{An example of the LE background estimation (ObsID: P030129310101). Top: spectrum of a blank sky observation (black) and the estimated background spectrum (red). Bottom: residuals in terms of errors ($\sigma$).}
		\label{fig:le_bkg_est}
\end{figure}

\begin{figure}[ht]%
	\centering
		\includegraphics[width=0.9\textwidth]{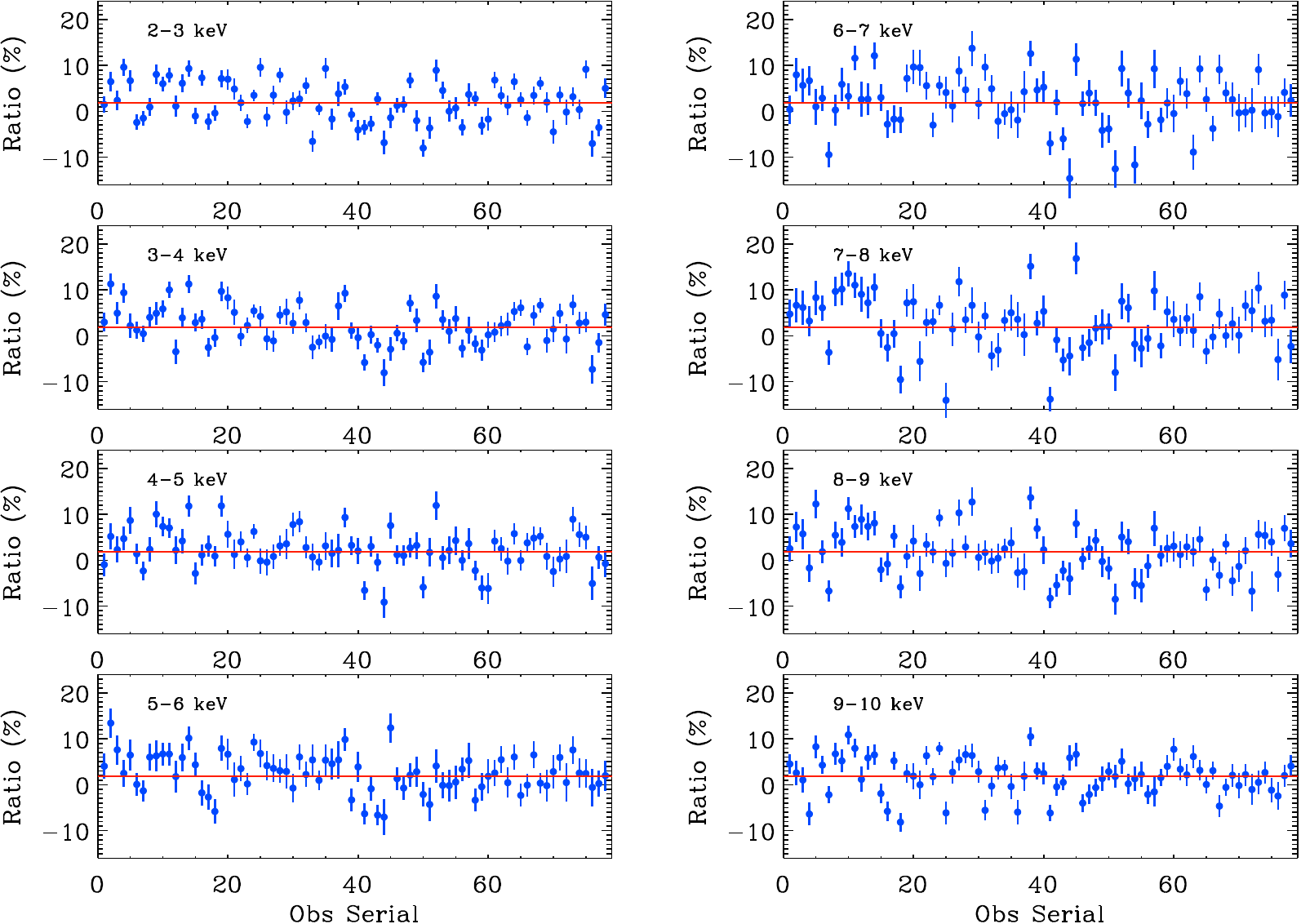}
		\caption{Deviations of the LE background estimation for eight energy bands in 4-th year.}\label{fig:le_bkg_est_eb}
\end{figure}

\begin{figure}[ht]%
	\centering
		\includegraphics[width=0.6\textwidth]{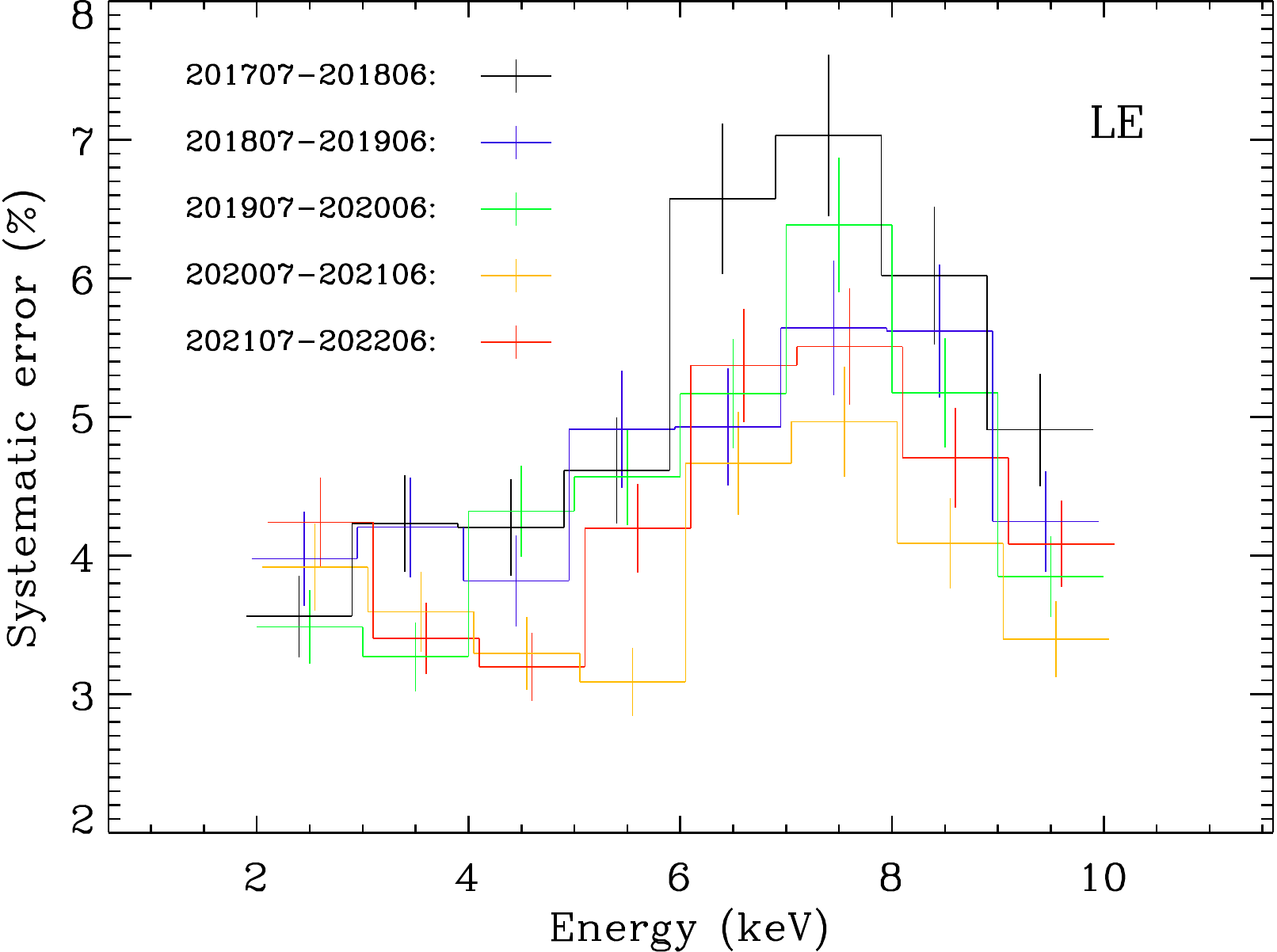}
		\caption{Systematic errors of the LE background model for every year after the launch of \textit{Insight}-HXMT.}\label{fig:le_err_sys}
\end{figure}

The validity of the background model is investigated, as it is critical for scientific analysis. With the same method in \cite{Liao2020_le}, we perform the background estimation for every blank sky observation. Fig.~\ref{fig:le_bkg_est} shows the background spectrum estimation (ObsID: P030129310101) as an example. 
For each year, the parameters of background model are updated to maintain the accuracy of the background estimate and then the systematic error of the background model is investigated. Fig.~\ref{fig:le_bkg_est_eb} shows the deviations of the LE background estimation in eight energy bands in the fourth year as an example.
The systematic errors of different energy bands between 2 to 10~keV in each year are shown in Fig~\ref{fig:le_err_sys}, and the results show that the systematic error does not change significantly compared to the first two years since the launch of \textit{Insight}-HXMT, i.e., the background model is stable and can give accurate background estimate. However, as the data around 1.5~keV is often affected by the electronic noise, the detection threshold is adjusted upward, thus only the systematic errors above 2~keV are given in this paper.

\section{Background of the Medium-Energy Telescope}\label{me_bkg}
\begin{figure}[ht]%
	\centering
		\includegraphics[width=0.6\textwidth]{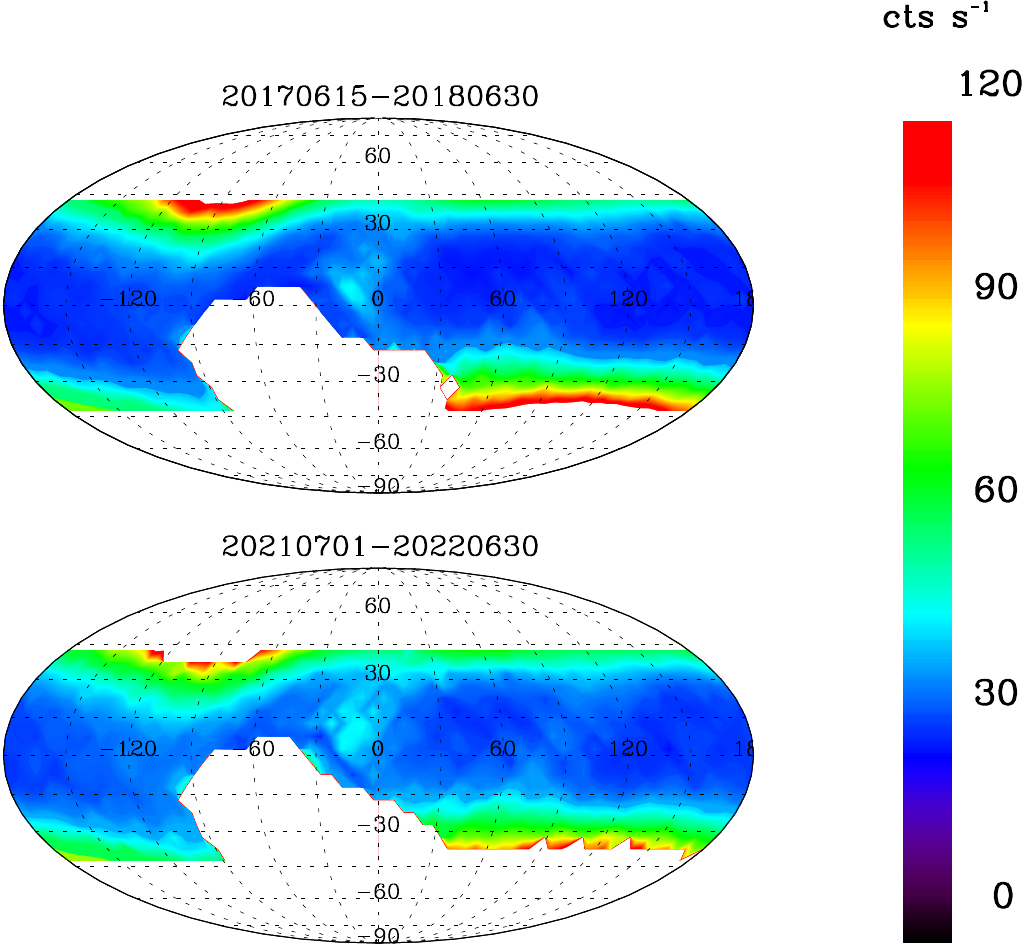}
		\caption{Geographical distributions of the background of ME small FoV detectors in the first and fifth years since \textit{Insight}-HXMT operation in orbit.}\label{fig:me_map}
\end{figure}

As shown in Table~\ref{tab:para}, ME is a collimating telescope sensitive in 5–40~keV with a total geometrical area of 952~$\rm cm^2$. It consists of 54 sensors in three boxes, and each sensor handles 32 SI-PIN pixels. For each box, there are 15 sensors with small FoV collimators and one sensor with the blocked FoV collimator that is used for background estimates.
The ME background characteristics have some similarities to those of LE in high energy band, especially the light curve feature and the geographical distribution.
However, the portion of each background component is very different, and the particle background is dominant in the whole energy band \citep{Guo2020,ZhangJ2020}.
Fig.~\ref{fig:me_map} shows a comparison of the ME background geographical distribution in the first and fifth years, respectively. It can be seen that the ME background in the fifth year is slightly higher than that in the first year.
In the region near SAA ($330^\circ<lon<360^\circ$, $0^\circ<lat<30^\circ$), the background is significantly higher than the most regions with similar latitudes.
This indicates that as the satellite passes through the high particle flux region (e.g., SAA), the ME background will rise firstly and then decay with time, i.e., the ME background has a delayed component.
Although the ME delayed background is relatively insignificant, it results in long time scale evolution of the ME background.
In order to improve the accuracy of the background estimation, the parameters of the background model should be given for every year.

\subsection{Observational characteristic and long-term evolution of the ME background}
\label{me_obs_char}

\begin{figure}[ht]%
	\centering
		\includegraphics[width=0.6\textwidth]{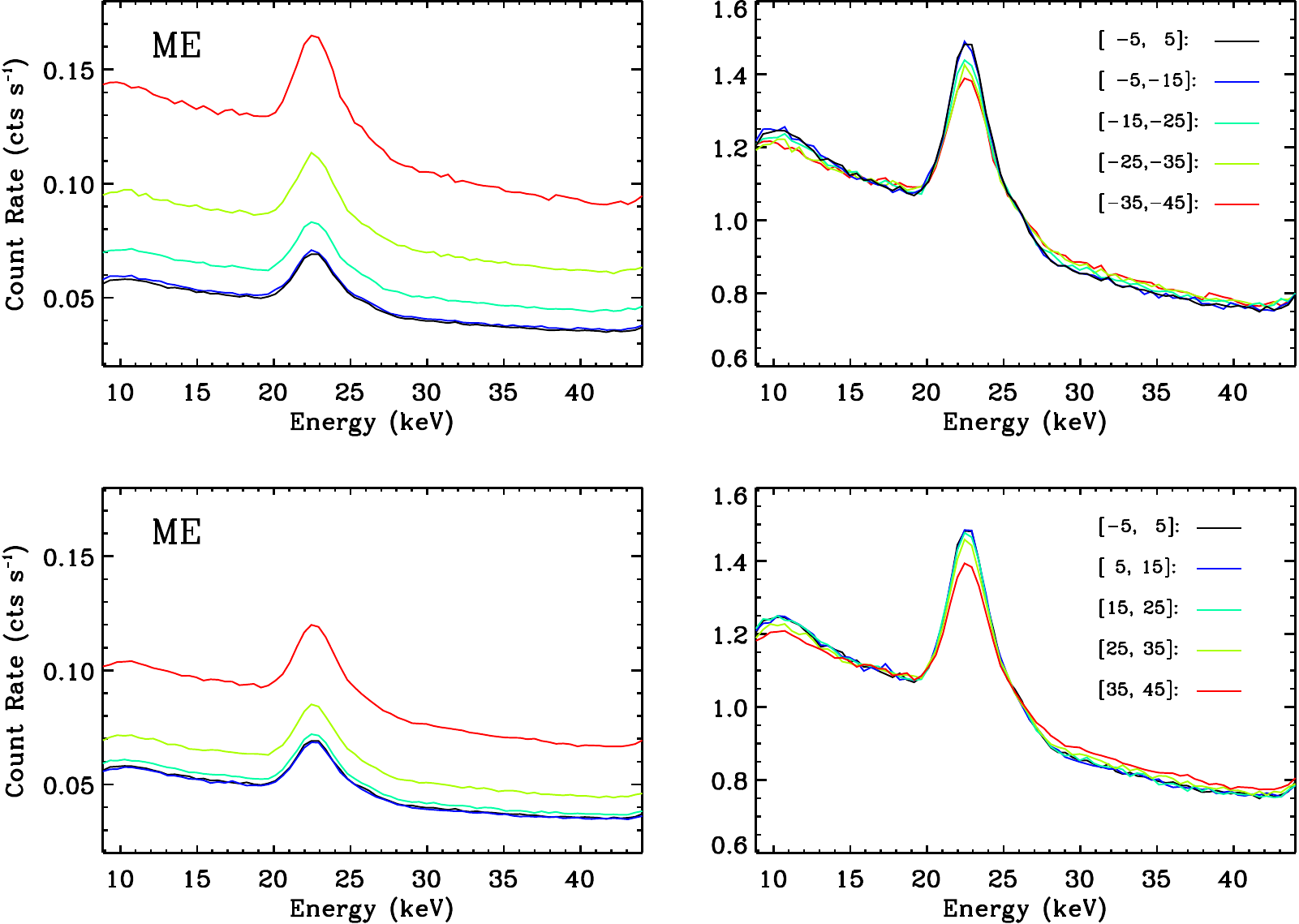}
		\caption{Spectra of the ME small FoV detectors in different geographical latitude ranges.}\label{fig:me_spec_geo}
\end{figure}

Fig.~\ref{fig:me_spec_geo} shows the spectra with different geographical latitude ranges. It can be seen that the intensity varies greatly, but the spectral shape keeps almost the same. Thus the evolution of the ME background, especially the intensity of the silver line, shall be carefully addressed in order to ensure the accuracy of the background model.

\begin{figure}[ht]%
	\centering
		\includegraphics[width=0.6\textwidth]{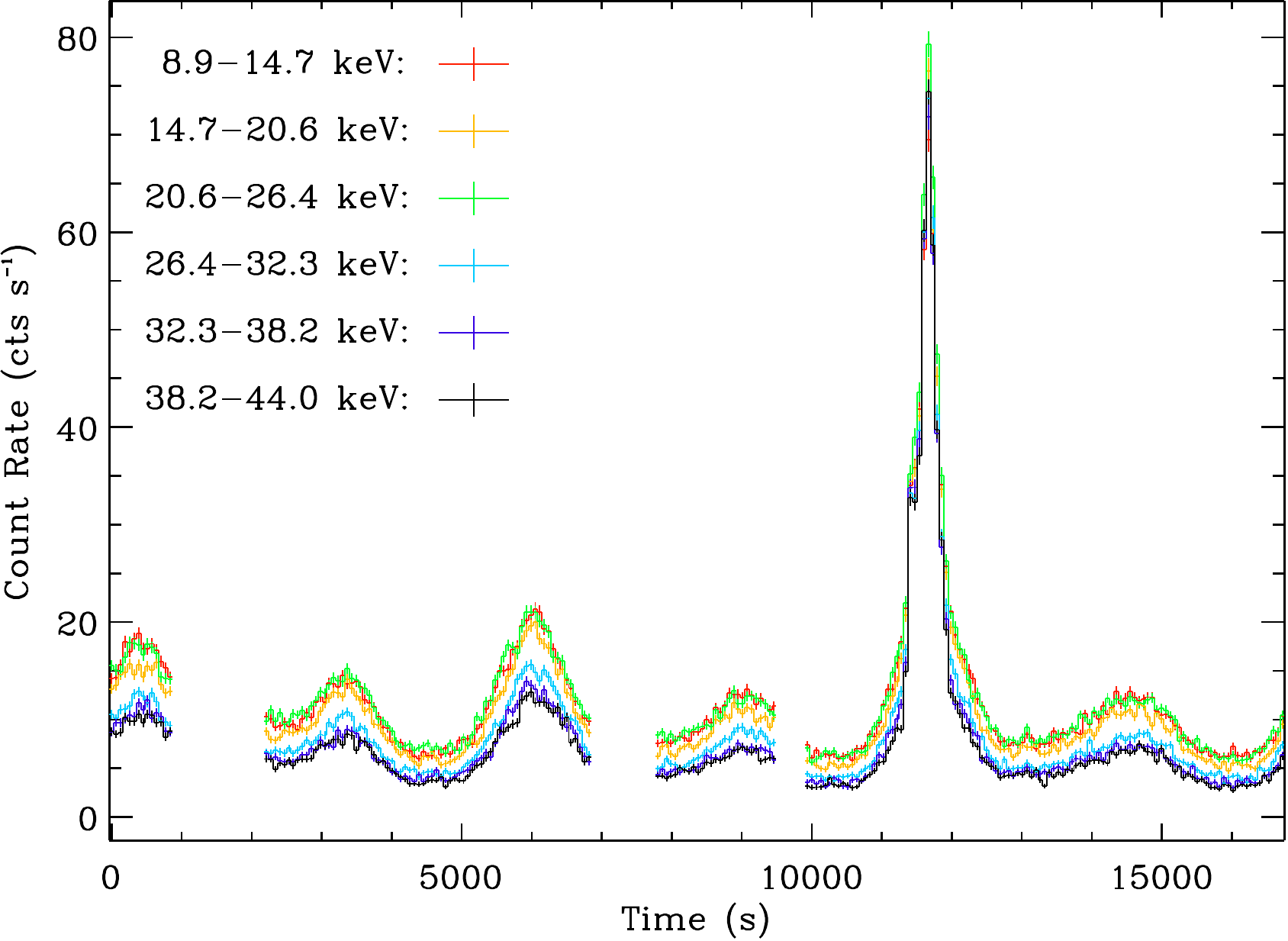}
		\caption{Light curves of the background observation by the ME small FoV detectors in six energy bands (ObsID: P040129309401, T0=2022-06-18T07:15:53.5).}\label{fig:me_lc}
\end{figure}

The ME background light curve exhibits significant orbital modulation (Fig.~\ref{fig:me_lc}).
A clear anti-correlation between ME background and geomagnetic cut-off rigidity has been shown in \cite{Guo2020}. There is an obvious peak in the light curve that is caused by particle events and is usually present in the high latitude region; accordingly, the corresponding time is excluded from GTI.

\begin{figure}[ht]%
	\centering
		\includegraphics[scale=0.4]{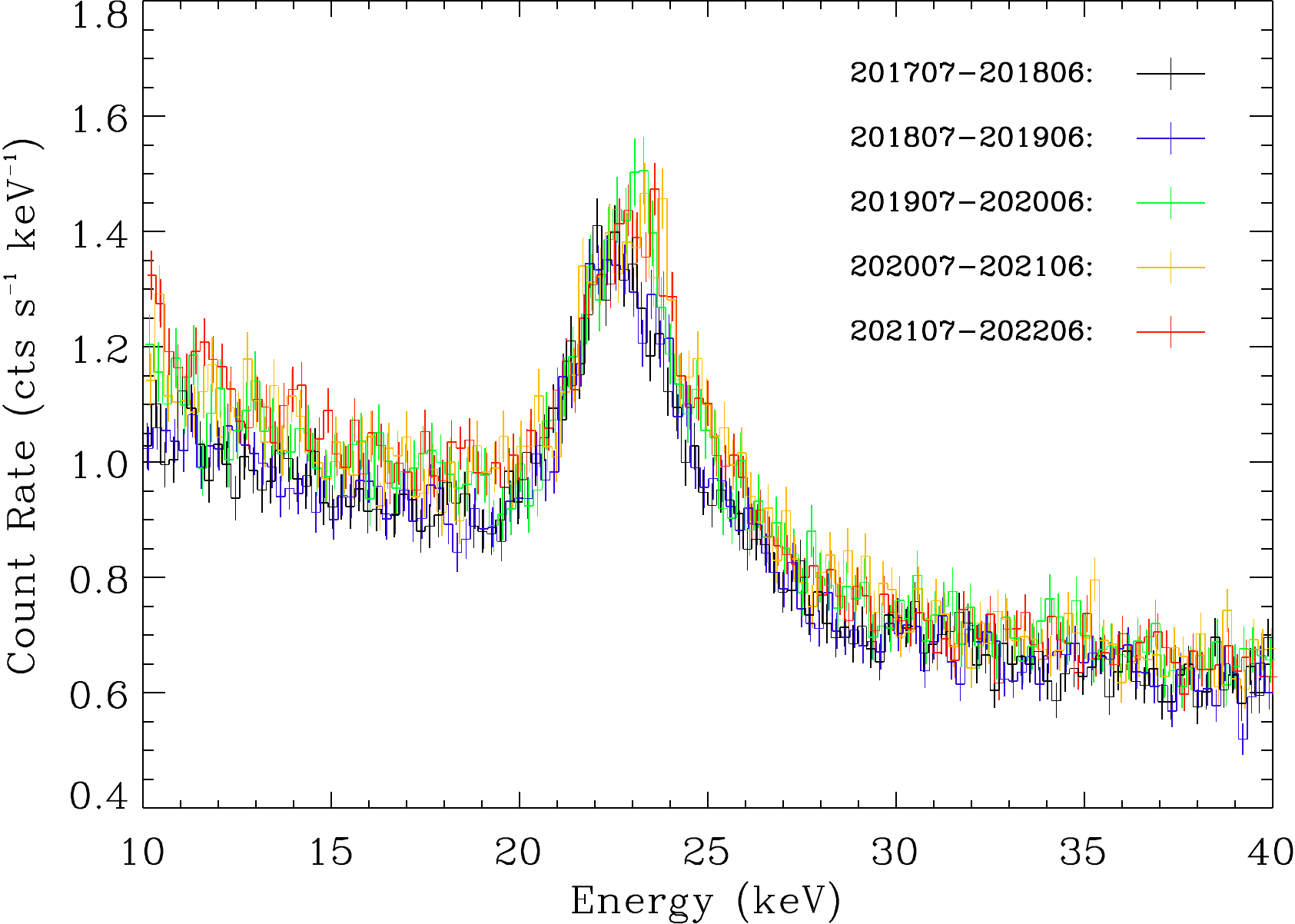}
		\caption{Background spectra of the ME small FoV detectors in every year since \textit{Insight}-HXMT operation in orbit.}\label{fig:me_small_spec}
\end{figure}

\begin{figure}[ht]%
	\centering
		\includegraphics[scale=0.4]{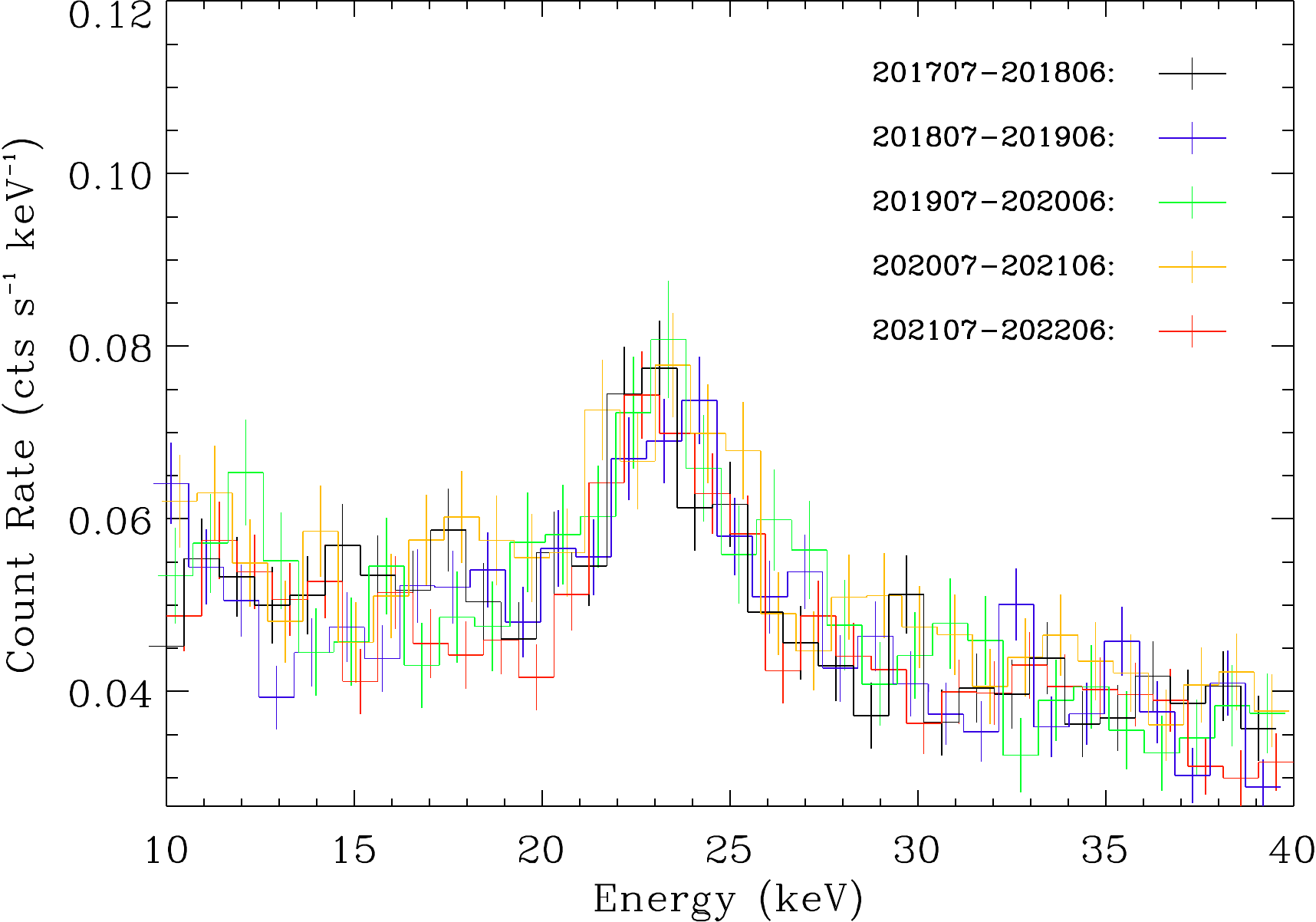}
		\caption{Same as Fig.~\ref{fig:me_small_spec} but for the blocked FoV detectors.}\label{fig:me_blind_spec}
\end{figure}

Fig.~\ref{fig:me_small_spec} shows the evolution of the background spectra of the small FoV detectors in the geographical region ($115^\circ<lon<125^\circ$, $-5^\circ<lat<5^\circ$). 
It can be seen that the ME background level increases with the increasing in-orbit operation as a cumulative effect of the weak delayed component. Since the delayed background increases with decreasing energy roughly \citep{ZhangJ2020}, the evolution of ME background in low energy band is more significant than that in high energy band. 
It is worth noting that the low-energy noise distributions have become wider as a result of the irradiation damage for some pixels, thus the spectrum below 11 keV for the fifth year has increased significantly.
The centers of the silver lines have also shifted over time, indicating a moderate change in the Energy-Channel relationship.
The spectral evolution of the blocked FoV detector is shown in Fig.~\ref{fig:me_blind_spec}. The positions of the silver lines do not shift significantly, which indicate that the blocked FoV detectors suffer less from radiation damage than the small FoV detectors.

\subsection{Validity of the ME background model}\label{me_bkg_model}
\cite{Guo2020} has constructed the background model and the corresponding database. Since the ME background spectral shape has a non-negligible change with the geographical location compared with LE, the average background of each detector at each geographical location must be considered in the background model.
In each background estimate, we first use the database to obtain the primary prediction spectra of the small FoV and blocked FoV detector in each geographical location the satellite has passed by, and then use the observation of the blocked FoV detector to make a further correction. The ME database produces the background spectra with time-averaged normalization for each individual geographical site, and the contemporary particle intensity can be determined by the blocked FoV detector and then used to correct the background model.

\begin{figure}[ht]%
	\centering
		\includegraphics[width=0.6\textwidth]{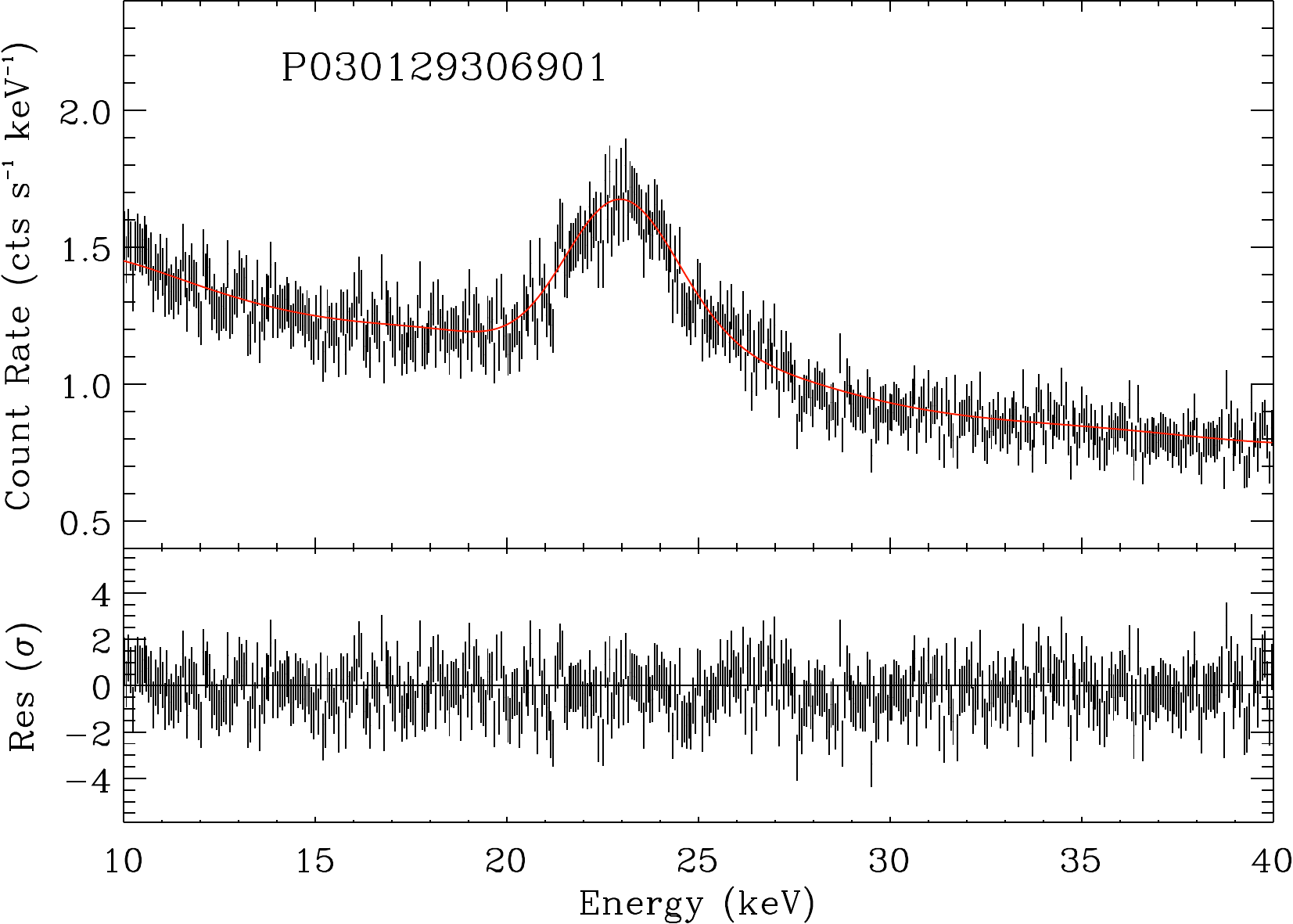}
		\caption{An example of the ME background estimation (ObsID: P030129306901). Top: spectrum of a blank sky observation (black points) and the estimated background spectrum (red line). Bottom: residuals in terms of errors ($\sigma$).}
		\label{fig:me_bkg_est}
\end{figure}

\begin{figure}[ht]%
	\centering
		\includegraphics[width=0.9\textwidth]{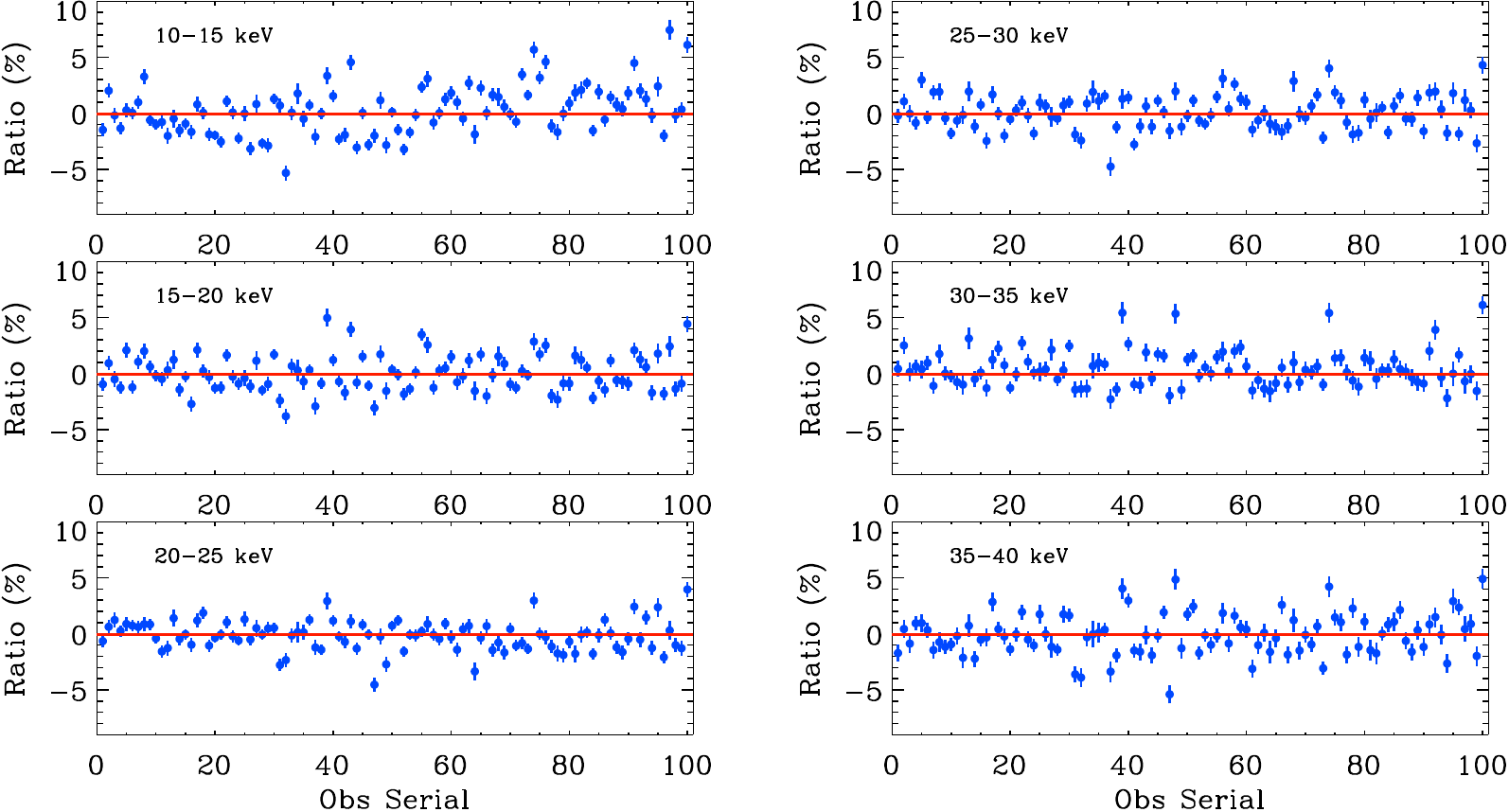}
		\caption{Deviations of the ME background estimation for six energy bands in the fourth year.}\label{fig:me_bkg_est_eb}
\end{figure}

\begin{figure}[ht]%
	\centering
		\includegraphics[scale=0.4]{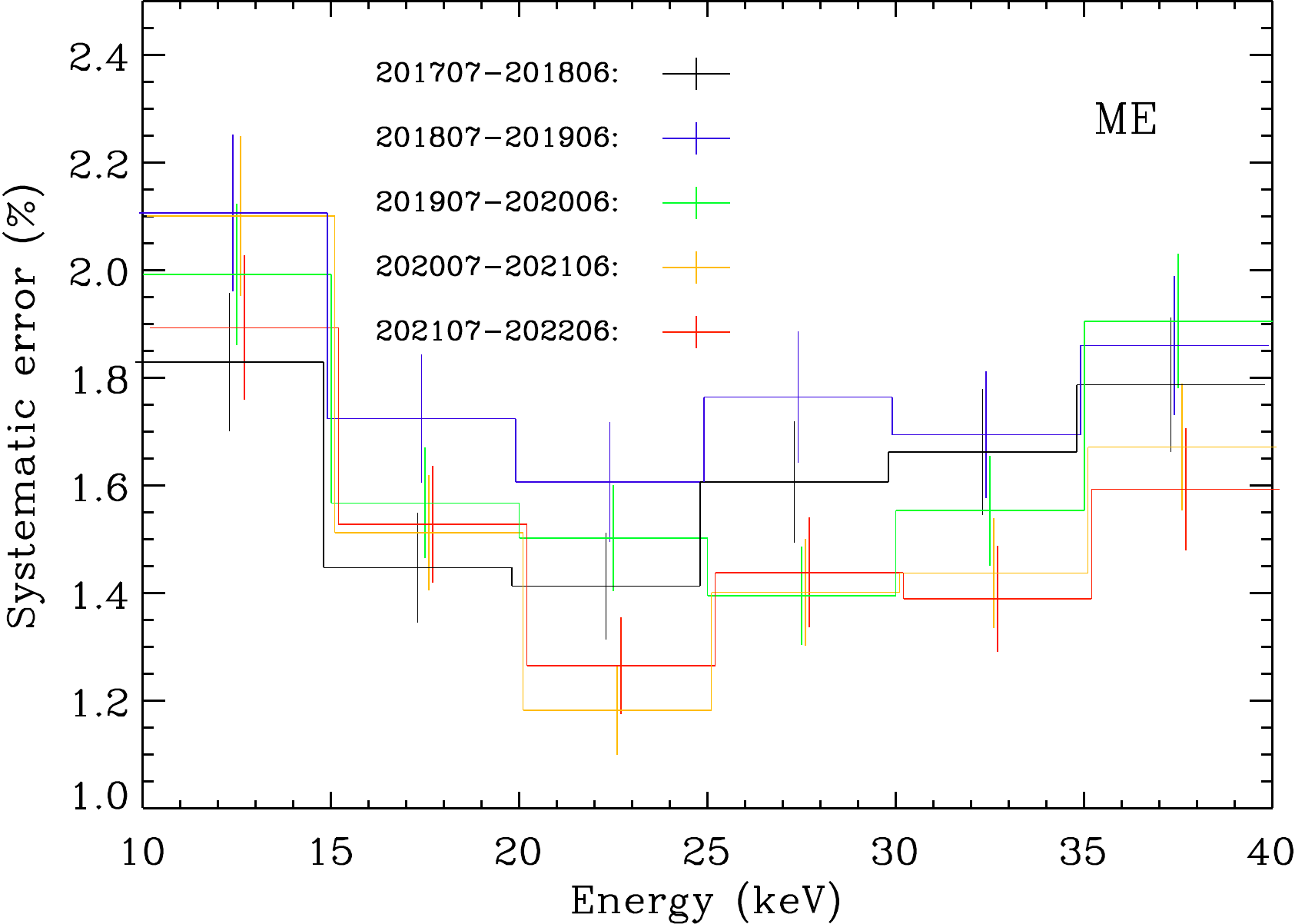}
		\caption{Systematic errors of the ME background model in every year.}\label{fig:me_err_sys}
\end{figure}

In each year, the backgrounds of all blank sky observations are estimated by the background model with the model parameters in the corresponding year. Fig.~\ref{fig:me_bkg_est} is an example of the ME background estimation to an blank sky observation (ObsID: P030129306901). A statistical analysis with the method in \cite{Guo2020} is performed to obtain the systematic error of the background estimation of each energy band. Fig.~\ref{fig:me_bkg_est_eb} shows the deviations of the ME background estimation for six energy bands in the fourth year.
Fig.~\ref{fig:me_err_sys} shows the systematic errors for six energy bands over the five years. The results show that the systematic error has no significant increasing trend over the five years. It can be seen that the systematic errors are relatively large in 10--15~keV with a mean value of $\sim2\%$, and the systematic errors in 10--40~keV are $\sim1.6\%$, indicating that the ME background model is still reliable.
It is worth mentioning that the detection below 10~keV is affected significantly by the electronic noise, so the current reliable energy range of ME begins from 10~keV.

\section{Background of the High-Energy Telescope}
\label{he_bkg}
HE has 18 NaI(Tl)/CsI(Na) phoswich detectors that are surrounded by 18 anti-coincidence detectors (ACDs) for active background shielding. Among the 18 detectors, 15 of which have a small FoV, two have a large FoV and one has a blocked FoV for background estimate.
The on-ground simulations have shown that the NaI and CsI crystals can be activated by the charged particles around the orbit of \textit{Insight}-HXMT, and the radioactive decay of the activated crystals is responsible for most of the HE background. As the satellite operates in orbit continuously, the crystals in HE detectors continue to be activated. After a significant rising in the first year, the rising trend of the HE background has gradually slowed down \citep{LiG2009,XieF2015}.

\subsection{Observational characteristic and long-term evolution of the HE background}\label{he_obs_char}

\begin{figure}[ht]%
	\centering
		\subfloat[\label{fig:he_map_a}]{\includegraphics[scale=0.5]{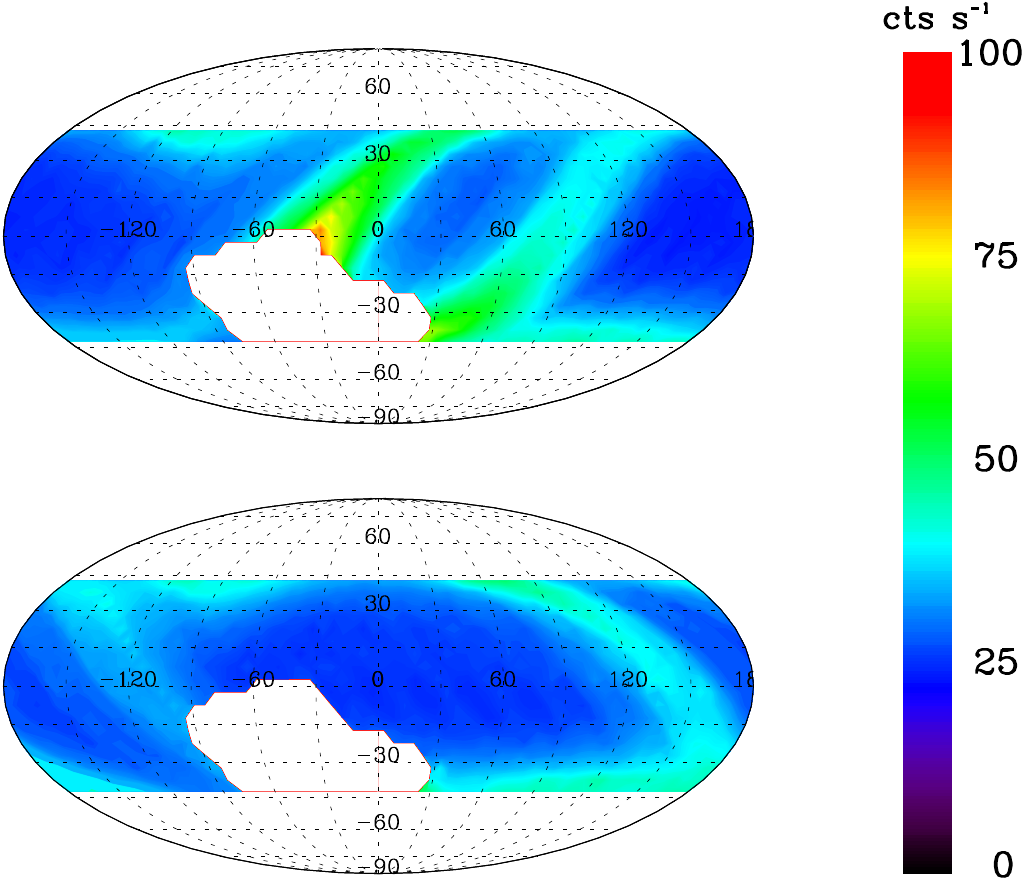}}
		\hspace{.26in}
		\subfloat[\label{fig:he_map_b}]{\includegraphics[scale=0.5]{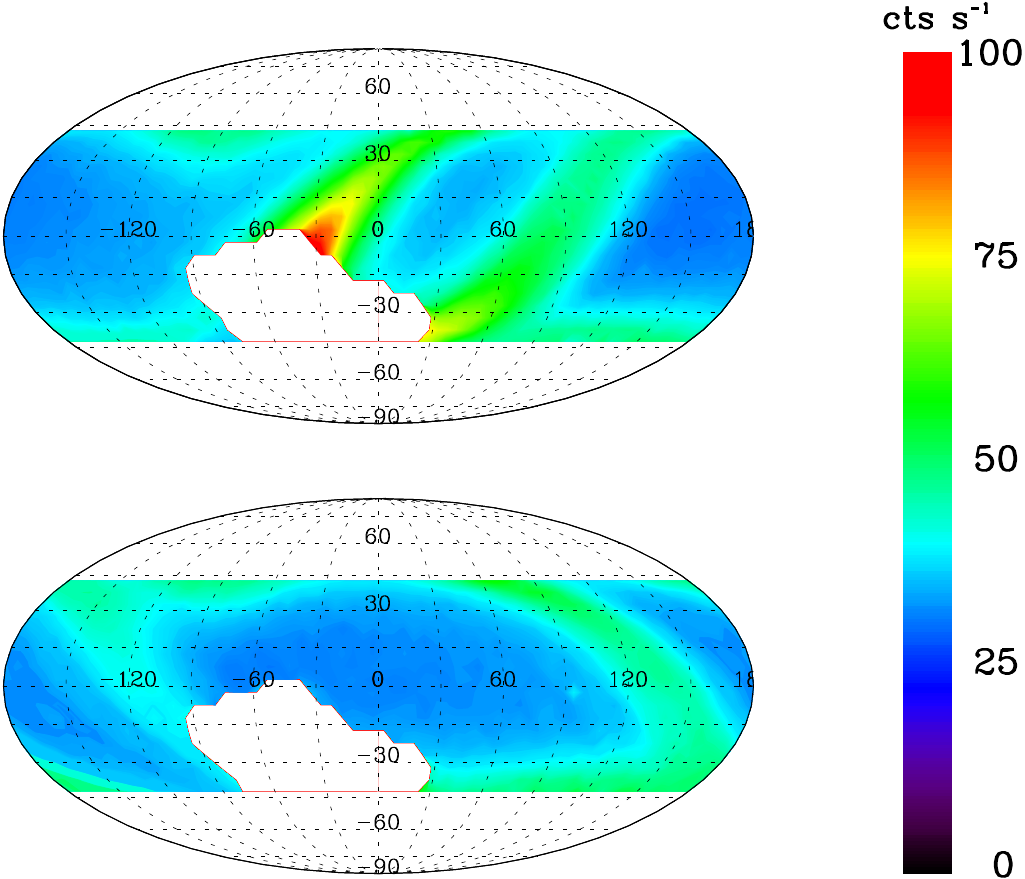}}
		\caption{Panel (a): Geographical distributions (top: ascending orbital phase, bottom: descending orbital phase)
		of the HE background intensity (DetID$=0$ \& $> 30~{\rm keV}$) in the first year. 
		Panel(b): same as the left panel but for the fifth year.}\label{fig:he_map}
\end{figure}

The graphical distributions of the HE background for the first and fifth years are shown in Fig.~\ref{fig:he_map}. It can be seen that the distribution has little difference, but the background count rate in the fifth year is significantly higher than that in the first year. Unlike the LE and ME backgrounds that are dominated by the prompt components, HE background is dominated by the time-delayed components due to the activation of the crystals by charged particles.
Consequently, the background is very different in the ascending and descending orbit phases even for the same geographical location.

\begin{figure*}[ht]
	\centering
		\subfloat[\label{fig:he_spec_ascend_0}]{\includegraphics[scale=0.3]{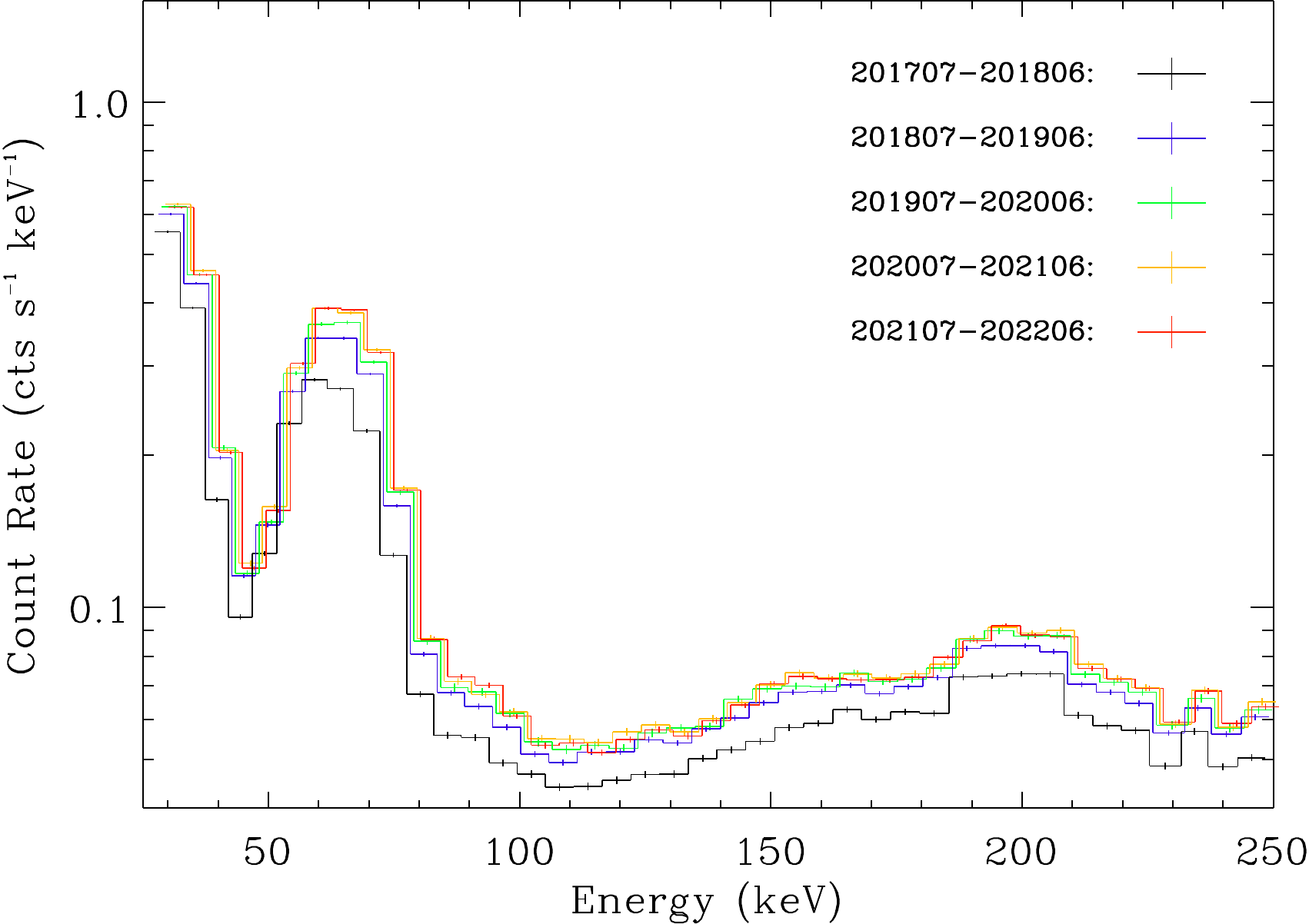}}
		\subfloat[\label{fig:he_spec_descend_0}]{\includegraphics[scale=0.3]{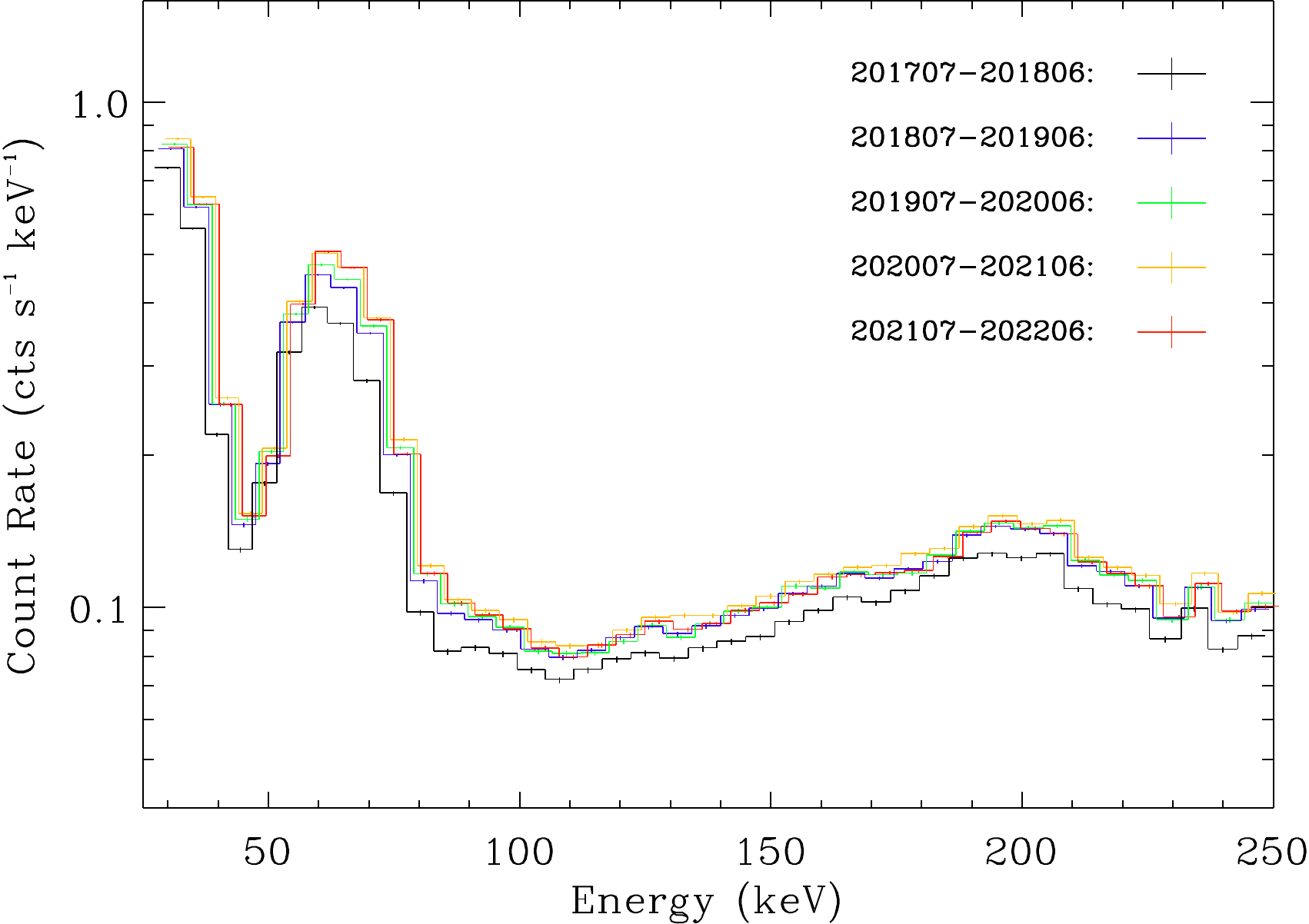}}
	\\
		\subfloat[\label{fig:he_spec_ascend_1}]{\includegraphics[scale=0.3]{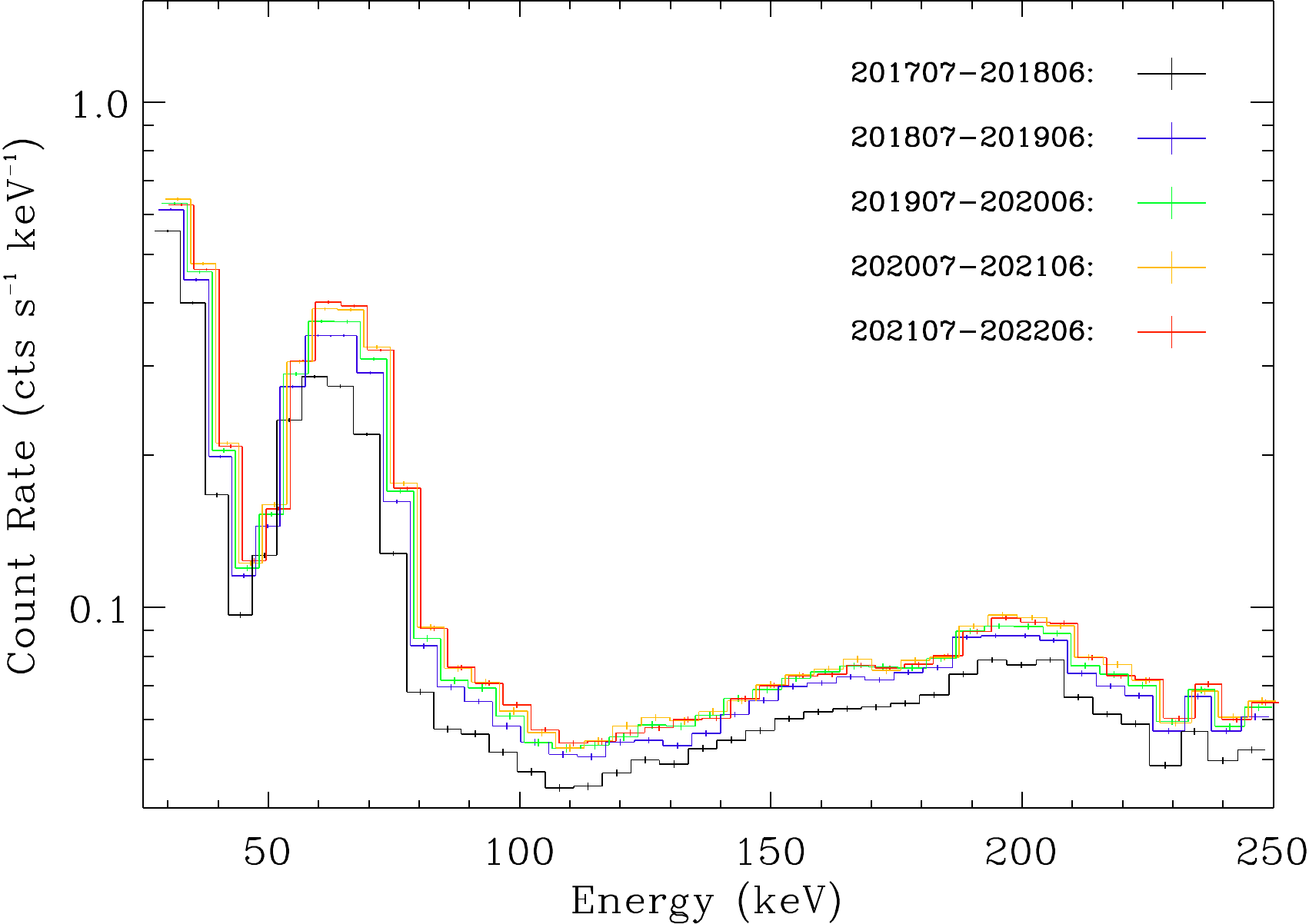}}
		\subfloat[\label{fig:he_spec_descend_1}]{\includegraphics[scale=0.3]{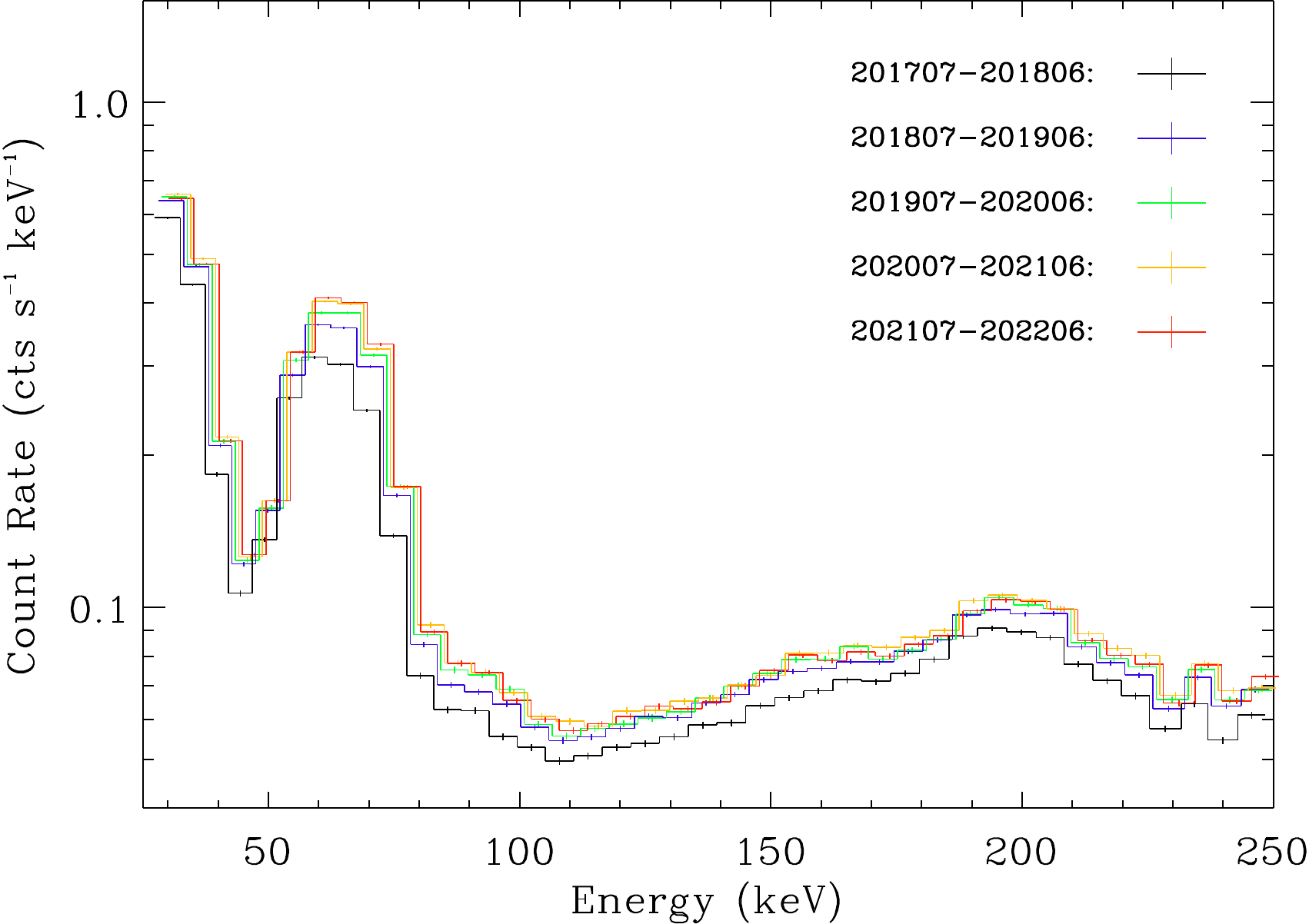}}
	\\
		\subfloat[\label{fig:he_spec_ascend_2}]{\includegraphics[scale=0.3]{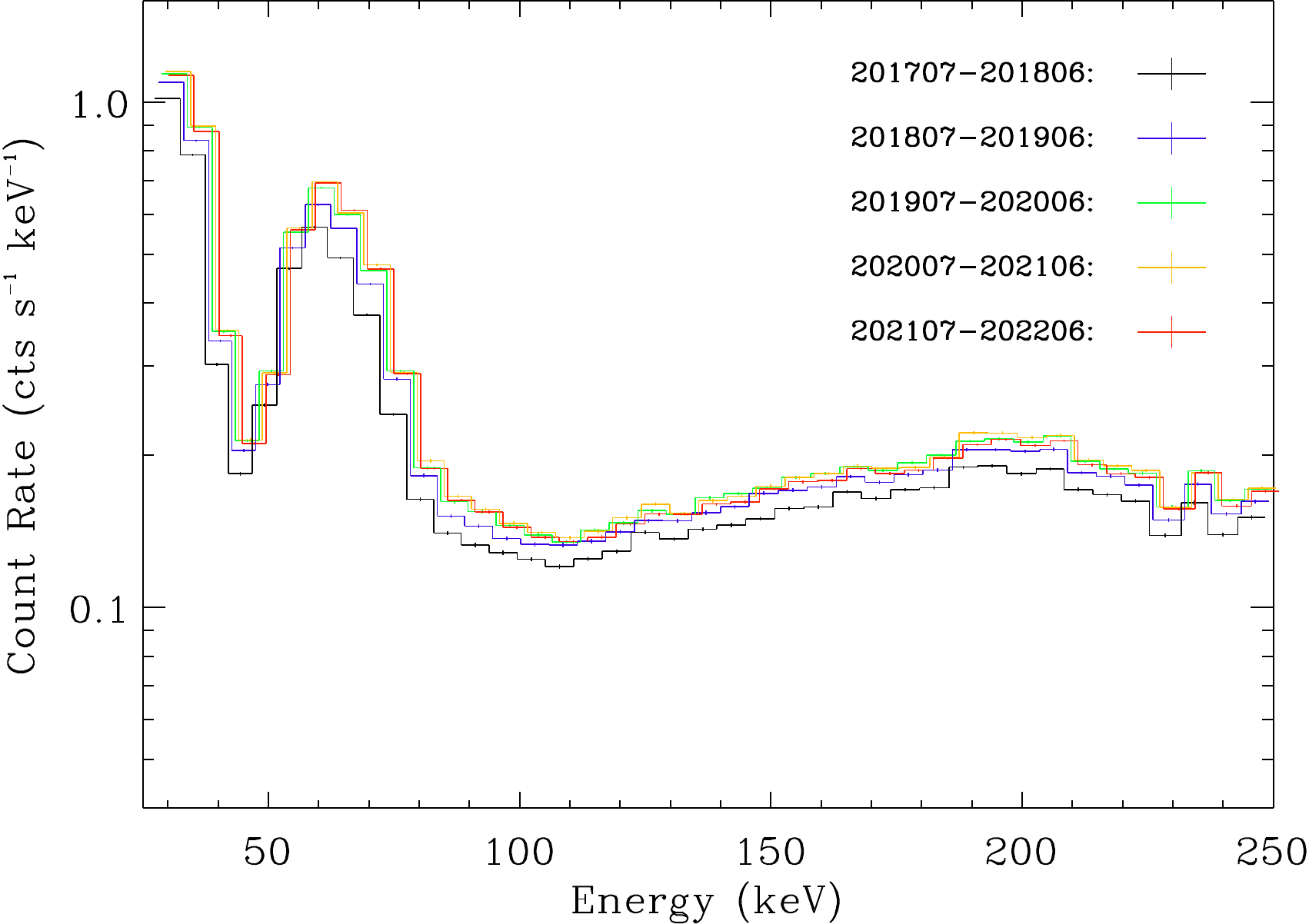}}
		\subfloat[\label{fig:he_spec_descend_2}]{\includegraphics[scale=0.3]{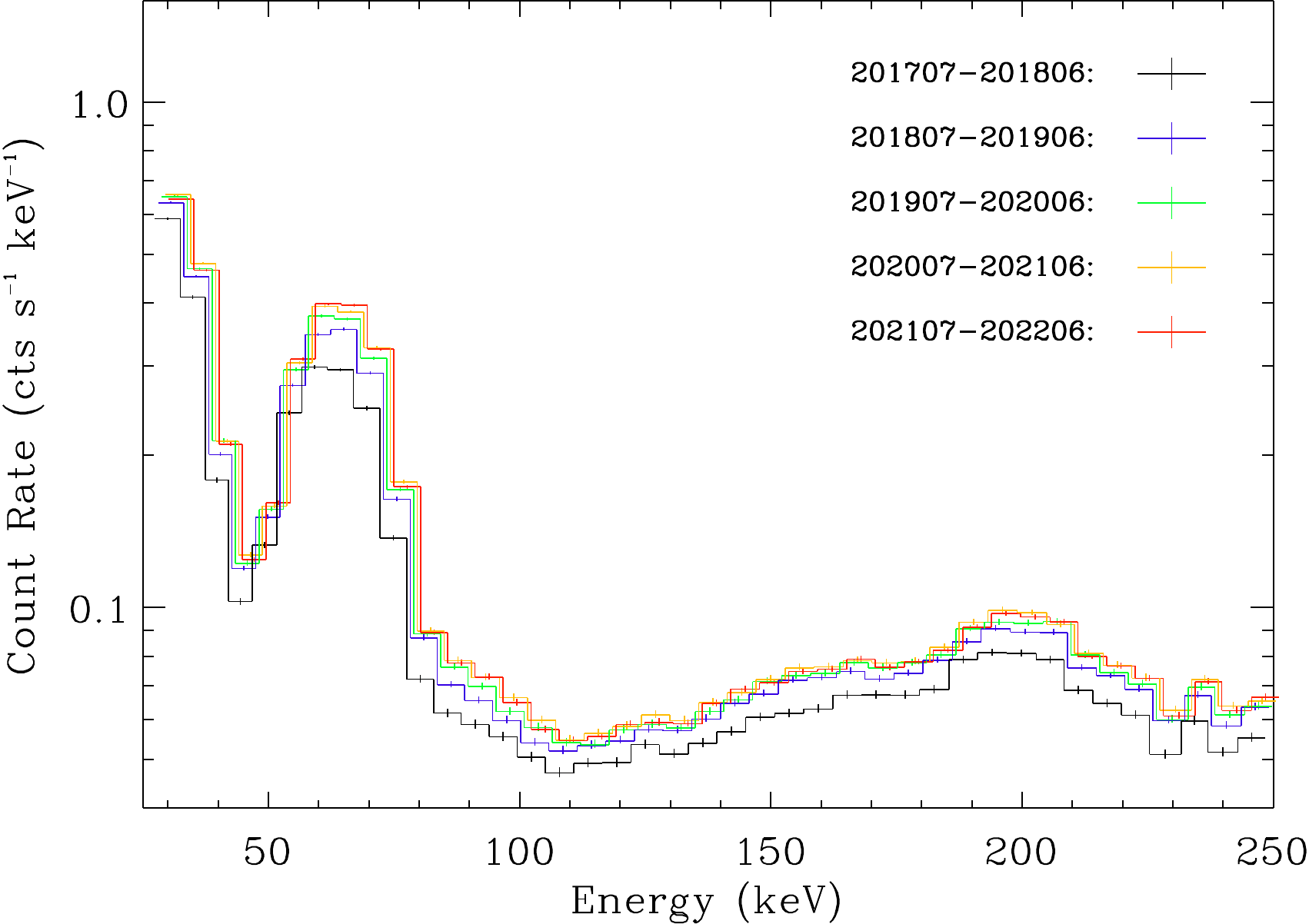}}
	\caption{Spectra of the HE background (${\rm DetID}=0$ \& $> 30~{\rm keV}$) in every year with different orbital phases and geographical locations.
		 	Left: Panel (a), (c) and (e) are these for $(lon,lat)$ = $(140^\circ,0^\circ)$, $(180^\circ,0^\circ)$ and $(345^\circ,15^\circ)$ in ascending orbital phase.
			Right: same as the left panels but for the descending orbital phase.}
	\label{fig:he_spec} 
\end{figure*}

Fig.~\ref{fig:he_spec} shows the spectra at geographical locations in ascending and descending orbit phases for every year.
The spectra in $(lon,lat)=(345^\circ,15^\circ)$ are shown in Panel (e) and (f).
In the ascending orbital phase, when the satellite passes through the SAA, the crystals of the detectors seriously activated do not have enough time to decay, and thus the background has a relatively high level. However, the background in descending orbit phase is lower. As it has been a long time since the last time the satellite passed through intense charged particle region and hence the background is dominated by the long-decay components.

The HE background spectra at different geographical locations show long-term evolution.
Compared with the results shown in other panels, the evolution shown in Panel (e) is less significant. This is because the satellite has just passed through the SAA, thus a large proportion of the background is contributed by the short-time scale component. Moreover, the intensity of SAA has not changed significantly over the five years (Fig.~\ref{fig:PM_map}).

As described in \cite{LiG2009} and \cite{XieF2015}, the spectra of HE background consist of various emission lines, which are induced by the interactions of the detectors with high-energy particles. It can be seen from Fig.~\ref{fig:he_spec} that the spectral shape is stable during the five years.

\begin{figure}[ht]%
	\centering
		\includegraphics[width=0.6\textwidth]{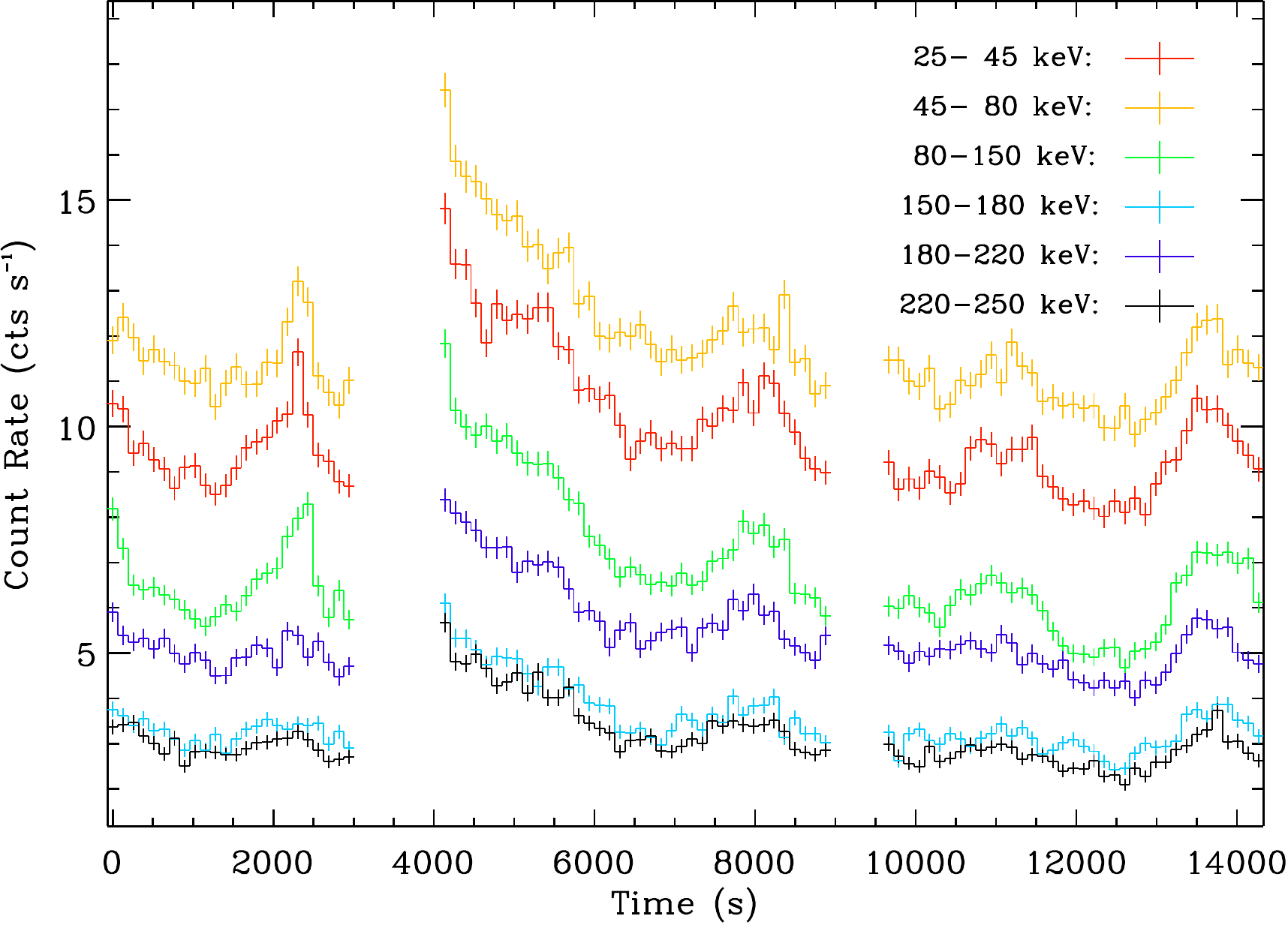}
		\caption{Light curves of the HE (DetID = 0) background observation in six energy bands (${\rm ObsID: P040129309001}$, T0 = 2022-05-22T18:15:38.5). 
		The gap is due to the protective shutdown of HE instrument when the satellite pass through SAA.}\label{fig:he_lc}
\end{figure}

Fig.~\ref{fig:he_lc} shows the light curves of HE observation of a blank sky region in six energy bands. For every energy band, the background rises to a high level when the satellite has just crossed SAA, then decays gradually and shows significant geomagnetic modulation.
There are also differences between the light curves in different energy bands, since the background is composed of numerous components characterized with different portions, spectral shapes and typical variable time scales.

\subsection{Validity of the HE background model}\label{he_bkg_model}
Based on the HE background characteristics, \cite{Liao2020_he} has made the HE background model. The principle is similar with that of ME but more complex. In order to obtain the background at any geographical location and any time, the empirical function with time as an independent variable is constructed to describe the long-term evolution of the HE background. Thus the preliminary estimation can be obtained with the orbital parameters and observation time, and some further correction will be performed with data of the blocked FoV detector.

\begin{figure}[ht]%
	\centering
		\includegraphics[width=0.6\textwidth]{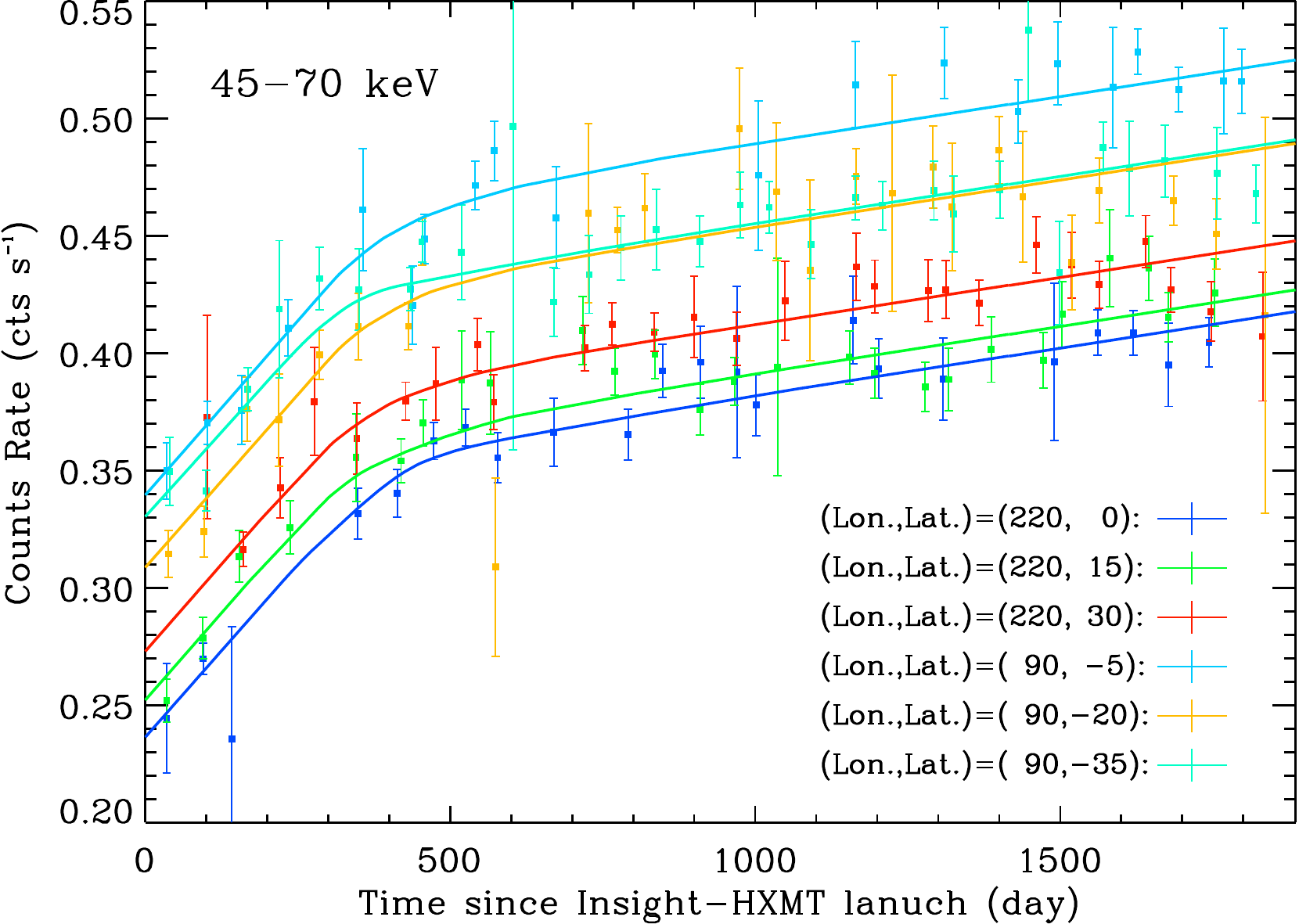}
		\caption{Long-term background evolution of the HE detector (DetID = 0) in 45--70~keV at six geographical locations.}\label{fig:he_long_term_lc}
\end{figure}

Therefore, HE background estimation is heavily dependent on the mathematical description of the background long-term evolution, i.e., the accuracy of the empirical function is critical to the background estimation. Fig.~\ref{fig:he_long_term_lc} shows the long-term evolution of background count rates at six different geographical locations in 45--70~keV. For each energy channel, the long-term evolution for each energy channel is described by a broken line with two slopes. The fitting curve is merged from the broken lines in this energy band with different break times, thus it shows a smooth transition without an obvious break.
It is worth noting that the broken line with two slopes is a function we chose to depict the long-term evolution of background count rate over time in an energy channel. 
Although other functions might be also acceptable, the selected broken-line shape could describe the observation data very well.
As predicted by the on-ground simulation \citep{LiG2009,XieF2015}, the activated isotopes lead to a rapid decay of the background rate after each SAA passage and a long-term accumulation as the in-orbit operation days increase. This accumulation rises rapidly in the initial epoch after launch and becomes slower after hundreds of days as the long lifetime isotopes are not dominant. This predicted long-term evolution is consistent with the observations shown in Fig.~\ref{fig:he_long_term_lc}.

\begin{figure}[ht]%
	\centering
		\includegraphics[width=0.6\textwidth]{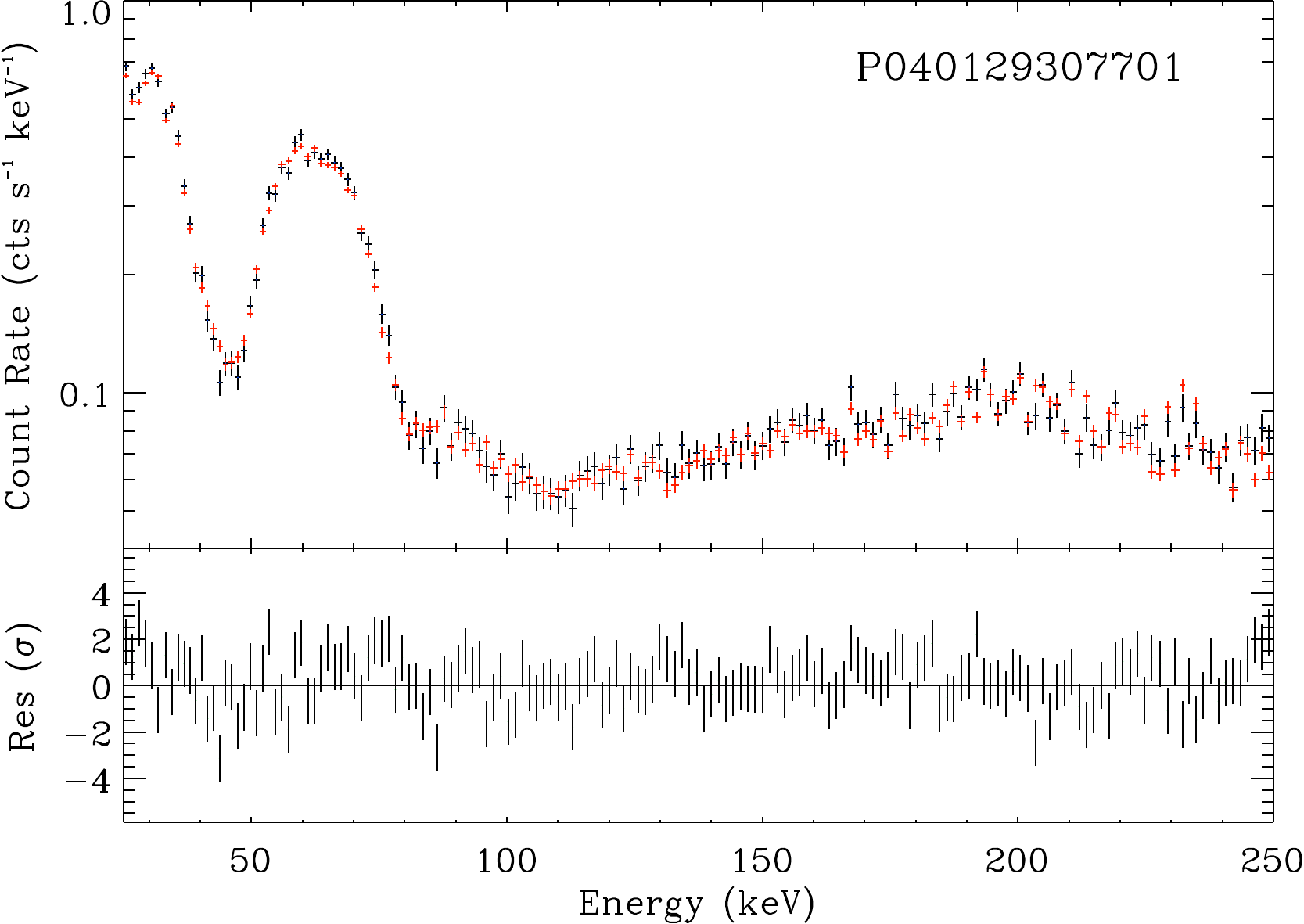}
		\caption{An example of the HE background spectrum estimation (ObsID: P040129307701). Top: spectrum of a blank sky observation (black) and the estimated background spectrum (red). Bottom: residuals in terms of errors ($\sigma$).}
		\label{fig:he_bkg_est}
\end{figure}

\begin{figure}[ht]%
	\centering
		\includegraphics[width=0.9\textwidth]{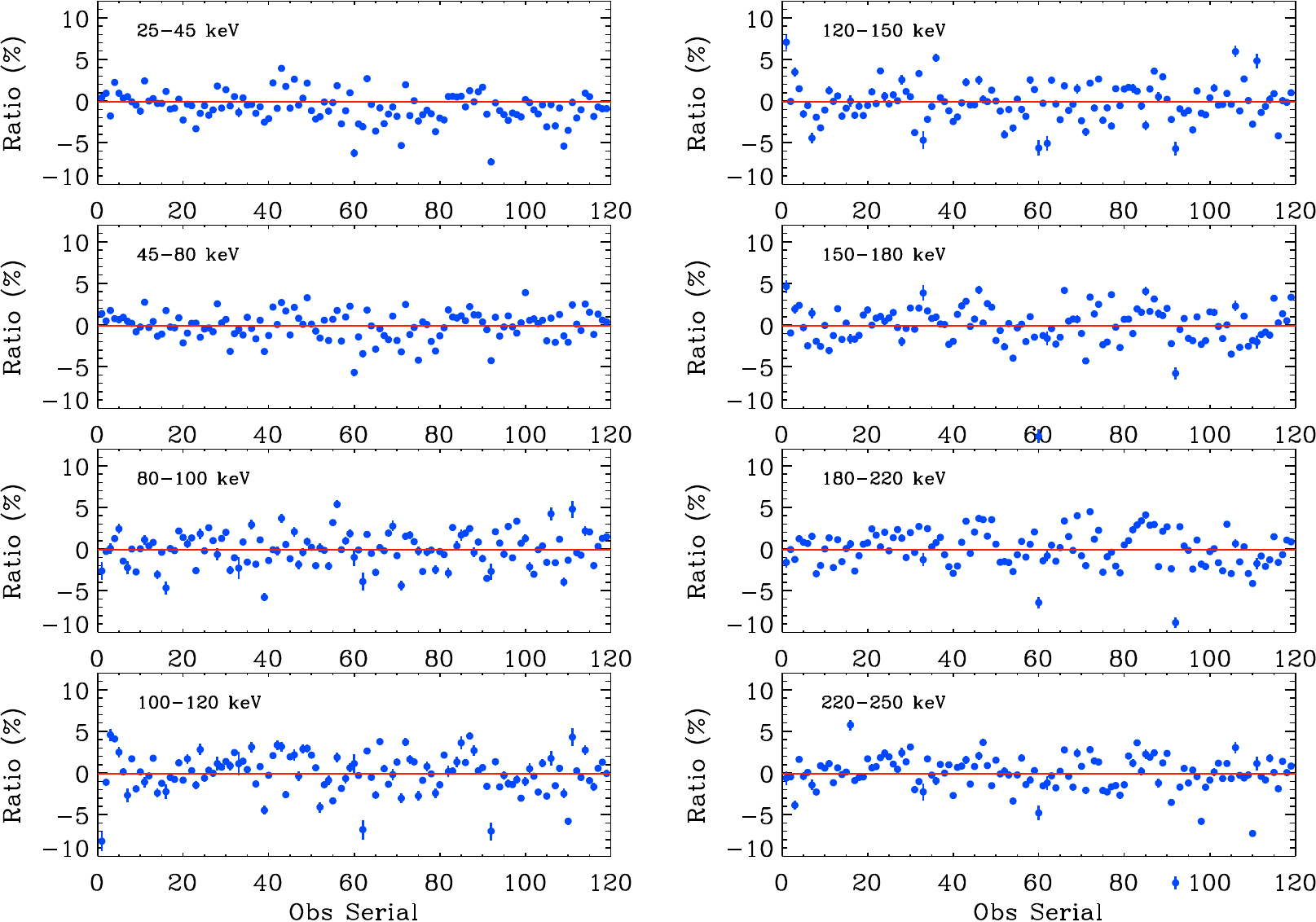}
		\caption{Deviations of the HE background estimation for eight energy bands in 5-th year.}\label{fig:he_bkg_est_eb}
\end{figure}

\begin{figure}[ht]%
	\centering
		\includegraphics[width=0.6\textwidth]{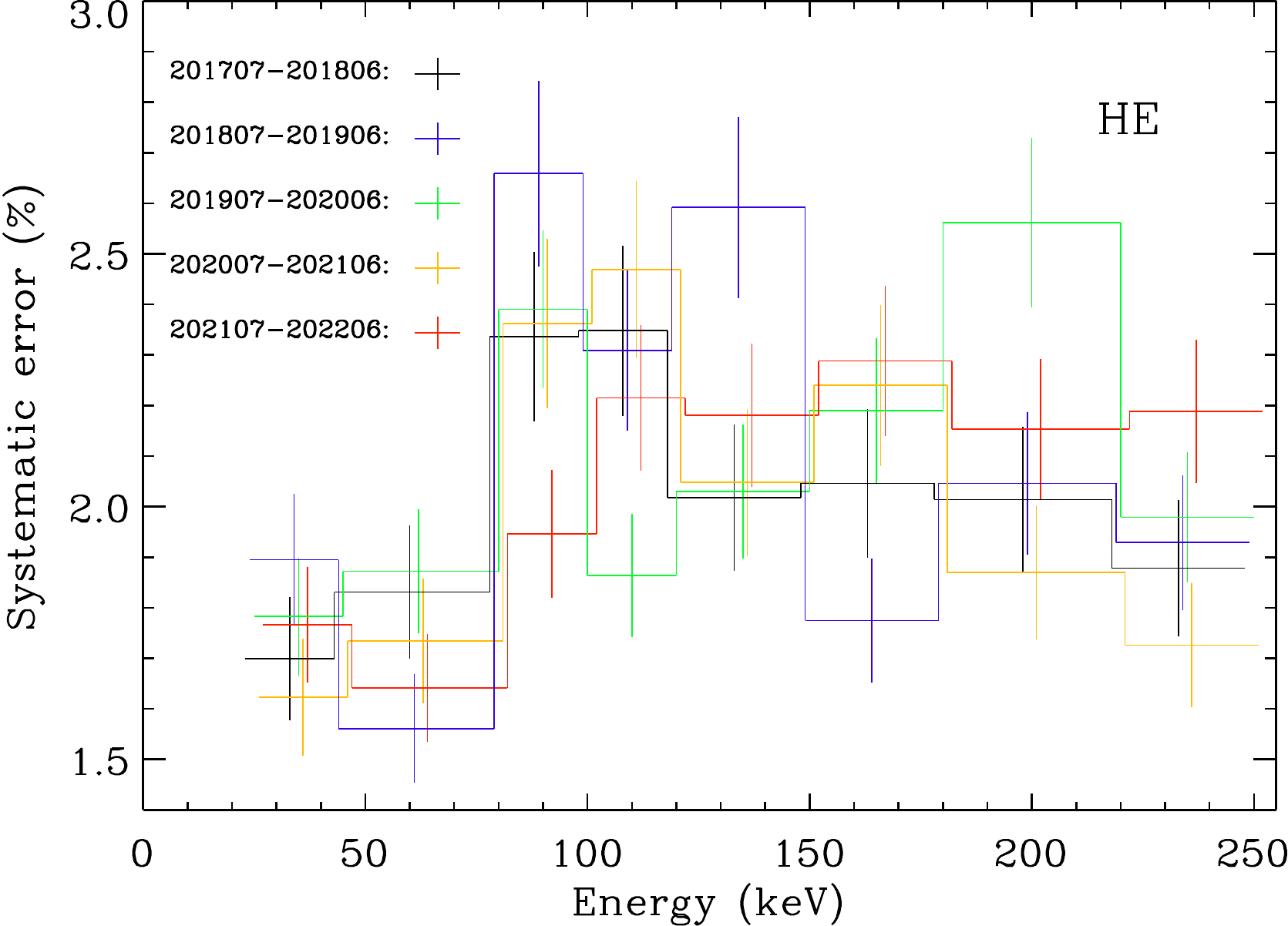}
		\caption{Systematic errors of the HE background model in 201707--202206.}
		\label{fig:he_err_sys}
\end{figure}

With the background model, the background estimate for all blank sky observations is preformed. Fig.~\ref{fig:he_bkg_est} is the background estimation to a blank sky observation (ObsID: P040129307701) shown as an example.
For each year, the deviations in eight energy bands between 25~keV and 250~keV (Fig~\ref{fig:he_bkg_est_eb} for the fifth year as an example) are obtained to
calculate the systematic errors.
Following the method in \cite{Liao2020_he}, the systematic errors can be obtained for each year (Fig.~\ref{fig:he_err_sys}). The results show that they are all less than 3\%, which is not much different from the results of the previous two years. So the HE background model is still effective.

\section{Summary and Conclusion}\label{summary_conclusion}

\textit{Insight}-HXMT has been operating successfully in orbit for five years. 
The backgrounds of the three telescopes exhibit different evolution trends compared with when \textit{Insight}-HXMT has just operated in orbit. 
All of the background behaviors are consistent with the expectations in design.

The evolution of the LE background is mainly caused by the subsequent operations to solve the problems brought by the LE irradiation damage, e.g., the lower limit of the detection threshold is adjusted higher to avoid the noise signal whose distribution becomes wider with increasing LE irradiation damage. Moreover, the continuous broadening of the emission lines in the background spectrum is the result of the decline of the LE energy resolution.
With the increasing  operation in orbit, the ME background level increases as a cumulative effect of the weak delayed component, and the variation of ME background in low energy band is more significant than that in high energy band. In addition, the ME background in low energy band  can also be affected by the low-energy noise of some pixels.
The crystals of HE detectors continue to be activated that is why the background intensity increases obviously with time. During the first five years, the increasing trend gradually slows down and shows a behavior similar to saturation. The background evolution at different energies is not consistent, which means that the shape of the background spectrum also evolves with time for a certain geographical location.

Although the time evolution characteristics of the LE and ME backgrounds are not significant, in order to maintain the accuracy of the background estimation, the background model parameters are updated in each year. For HE background model, the evolution has been taken into account from the beginning of the model construction.
The statistical analysis shows that the systematic errors of the three telescopes change little during the first five years of \textit{Insight}-HXMT operation, thus the background models are still effective and reliable.

As described in \cite{Liao2020_le}, the LE background model constructed with the blank sky observations can effectively estimate for both the particle background and the diffuse background caused by the CXB. Therefore, it can be used to the pointing observation with the target in high Galactic latitude ($\lvert{b}\rvert\gt10^\circ$). 
In order to estimate the diffuse background accurately in low Galactic latitude ($\lvert{b}\rvert\le 10^\circ$), the diffuse X-ray background in the Galactic Plane \citep{Jin2022} should be used in the LE background estimation.

It is worth noting that the current background models of the three telescopes rely heavily on the blocked FoV detectors. So the blocked FoV detectors are critical, especially for HE as it has only one blocked FoV detector. This is a potential hazard for background estimates because of insufficient robustness. Therefore, an alternative to background estimation that does not rely on blocked FoV detectors must be planned in advance, such as using ACD and PM as prompt particle monitors for background estimation of LE and ME. For HE, a parametric background model that does not rely on blocked FoV detectors has been built. By considering the physical factors that generate the HE background, a mathematical model considering these physical processes has been successfully constructed \citep{You2021}.

\bmhead{Acknowledgments}
This work made use of the data from \textit{Insight}-HXMT mission, a project funded by China National Space Administration (CNSA) and the Chinese Academy of Sciences (CAS). 
The authors thank supports from the National Key R\&D Program of China (2021YFA0718500), the National Natural Science Foundation of China under Grants Nos. U1838202, U1838201. This work was partially supported by International Partnership Program of Chinese Academy of Sciences (Grant No.113111KYSB20190020).

\section*{Declarations}
On behalf of all authors, the corresponding author states that there is no conflict of interest.


\begin{thebibliography}{99}

\bibitem[Alcaraz et al.(2000a)]{Alcaraz2000a}
Alcaraz, J., Alvisi, D., Alpat, B., et al. 2000, Phys. Lett. B, 472, 215

\bibitem[Alcaraz et al.(2000b)]{Alcaraz2000b}
Alcaraz, J., Alpat, B., Ambrosi, G., et al. 2000, Phys. Lett. B, 484, 10

\bibitem[Cao et al.(2020)]{CaoXL2020}
Cao, X., Jiang, W., Meng, B., et al.\ 2020, Science China Physics, Mechanics, and Astronomy, 63, 249504

\bibitem[Chen et al.(2020)]{ChenY2020}
Chen, Y., Cui, W., Li, W., et al.\ 2020, Science China Physics, Mechanics, and Astronomy, 63, 249505

\bibitem[Frontera et al.(1997a)]{Frontera1997a}
Frontera, F., Costa, E., dal Fiume, D., et al., 1997a. Astron. Astrophys. Suppl. Ser. 122, 357

\bibitem[Frontera et al.(1997b)]{Frontera1997b}
Frontera, F., Costa, E., dal Fiume, D., et al., 1997b. Proc. SPIE 3114, 206.

\bibitem[Guo et al.(2020)]{Guo2020}
Guo, C.-C., Liao, J.-Y., Zhang, S., et al.\ 2020, Journal of High Energy Astrophysics, 27, 44

\bibitem[Jin et al.(2022)]{Jin2022}
Jin, J., Liao, J.-Y., Wang, C., et al. 2022, ApJS, 260, 42

\bibitem[Li et al.(2009)]{LiG2009}
Li, G., Wu, M., Zhang, S., et al., 2009, Chin. Astron. Astrophys, 33, 333

\bibitem[Liao et al.(2020a)]{Liao2020_le}
Liao, J.-Y., Zhang, S., Chen, Y., et al.\ 2020, Journal of High Energy Astrophysics, 27, 24

\bibitem[Liao et al.(2020b)]{Liao2020_he}
Liao, J.-Y., Zhang, S., Lu, X.-F., et al.\ 2020, Journal of High Energy Astrophysics, 27, 14

\bibitem[Liu et al.(2020)]{LiuCZ2020}
Liu, C., Zhang, Y., Li, X., et al.\ 2020, Science China Physics, Mechanics, and Astronomy, 63, 249503

\bibitem[Lu et al.(2020)]{LuXF2020}
Lu, X., Liu, C., Li, X., et al.\ 2020, Journal of High Energy Astrophysics, 26, 77

\bibitem[Luo et al.(2020)]{Luo2020}
Luo, Q., Liao, J.-Y., Li, X.-F., et al. 2020, Journal of High Energy Astrophysics, 27, 1

\bibitem[Rothschild et al.(1998)]{Rothschild1998}
Rothschild, R. E., Blanco, P. R., Gruber, D. E., et al. 1998, ApJ, 496, 538

\bibitem[Sai et al.(2020)]{Sai2020}
Sai, N., Liao, J.-Y., Li, C.-K., et al. 2020, Journal of High Energy Astrophysics, 26, 1

\bibitem[Xie et al.(2015)]{XieF2015}
Xie, F., Zhang, J., Song, L.M., et al. 2015, Ap\&SS, 360, 47

\bibitem[You et al.(2021)]{You2021}
You, Y., Liao, J.-Y., Zhang, S.-N., et al. 2021, ApJS, 256, 47

\bibitem[Zhang et al.(2020)]{ZhangJ2020}
Zhang, J., Li, X.B., Ge, M.Y., et al. 2020, Ap\&SS, 365, 158

\bibitem[Zhang et al.(2020)]{ZSN2020}
Zhang, S.-N., Li, T.-P., Lu, F.-J., et al.\ 2020, Science China Physics, Mechanics, and Astronomy, 63, 249502


\end{thebibliography}
\end{document}